\documentclass[runningheads]{llncs}

\usepackage{times}
\usepackage{array}
\usepackage{epsfig}
\usepackage{graphicx}
\usepackage{amsmath}
\usepackage{amssymb}
\usepackage{subcaption}
\usepackage{float}
\restylefloat{table}

\usepackage{multirow}

\usepackage{dblfloatfix}

\usepackage{eccv}

\usepackage{eccvabbrv}

\usepackage{booktabs}

\usepackage[accsupp]{axessibility}  

\usepackage{hyperref}

\usepackage{orcidlink}

\begin{document}

\title{One-Shot Image Restoration}

\titlerunning{One-Shot Image Restoration}

\author{Deborah Pereg\inst{1,2}\orcidlink{0000-0002-2453-6577}}

\authorrunning{D.~Pereg}


\institute{Massachusetts Institute of Technology \and Harvard School of Engineering $\&$ Applied Sciences \\
Cambridge, MA, USA \\
\email{deborahp@mit.edu}}

\maketitle
\begin{abstract}
Image restoration, or inverse problems in image processing, is an extensively studied topic. In recent years supervised learning approaches have become a popular strategy attempting to tackle this task. Unfortunately, most supervised learning-based methods are highly demanding in terms of computational resources and training data (sample complexity). Moreover, trained models are sensitive to domain changes, such as varying acquisition systems, signal sampling rates, resolution and contrast.  
In this work, we try to answer a fundamental question: Can supervised learning models generalize well solely by learning from one image or even part of an image? If so, then what is the minimal amount of patches required to achieve acceptable generalization?
To this end, we focus on an efficient patch-based learning framework that requires a single image input-output pair for training. Experimental results demonstrate the applicability, robustness and computational efficiency of the proposed approach for supervised image deblurring and super resolution. Our results showcase significant improvement of learning models' sample efficiency, generalization and time complexity, that can hopefully inspire future deployment of the proposed  learning perspective to future real-time applications, and be applied to other signals and modalities.


\end{abstract}

\section{Introduction}\label{sec1}

In recent years, supervised learning methods have accomplished remarkable results across numerous fields of knowledge. However, improving sample efficiency and generalization, to enable efficient deployment and implementation of artificial intelligence (AI) algorithms in practical applications, still poses a significant challenge. Moreover, adaptation of supervised learning models to domain changes continues to present a challenging obstacle. In particular, deep neural networks (DNNs) designed for computational imaging tasks are susceptible to alternations in the physical parameters of the acquisition system, such as: sampling space, resolution, and contrast. Even within the same acquisition system, performance is known to notably degrade across datasets.

The majority of supervised learning methods concentrate on learning from thousands of examples, training that can last hours and sometimes weeks, requires expensive {GPU}s \cite{Alwosheel:2018}, and leaves a disturbing carbon footprint \cite{Strubell:2020}. Thus, the main objective of this research paper is to investigate the possibility of learning from a single-image example for image restoration tasks.

Evidence of training based on few examples can be found in other supervised learning frameworks, such as few-shot classification \cite{Snell:2017,Vinyals:2016, Miller:2000,Lake:2011}.
TinyML \cite{Lin:2020} aims to run deep learning models on low-power IoT devices with tight memory constraints. 
On-device training \cite{Lin:2022} could benefit from the ability to train with a single signal example, which requires substantially smaller memory, and avoids the need for expensive GPUs. Few-shot transfer learning has also become increasingly popular recently \cite{Huang:2022,Wu:2020}. 

Although, humans are known to be able to perform one-shot learning, 
the ability of supervised learning models to generalize well from a single example in low vision tasks in image processing has been a longstanding controversial statement \cite{Pereg:2022}. 
Zontak $\&$ Irani (2011) postulate that patches of the same image are internally repeated within that image, but unlikely to be found in other images.
Several directions for internal learning, i.e., zero-shot learning from a single input based on the principle of self-similarity were investigated lately \cite{Tirer:2023}. 
Shocher et al. (2018) \cite{Shocher:2018} train a small image-speciﬁc CNN at test time, on super resolution examples extracted solely from the input image itself, based on the property of internal recurrence of information inside a single image \cite{Irani:2011}. 

The principle of internal learning might appear as a contradiction to the proposed approach. Nevertheless,
in \cite{Pereg:2023A} we proved that based on information theoretic asymptotic equipartition property (AEP) \cite{ThomasCover:2006}, there is a relatively small set that can empirically represent the input-output data distribution for learning. Hence, in light of the theoretical analysis, it may be claimed that image patches that could reliably represent the typical set associated with natural images would suffice for good generalization. Note that using a substantially larger training data as commonly accepted does not contradict the AEP, since the realization of the sequences in the typical set would be expressed in the large training data. 

Previous works have demonstrated the ability to successfully generalize from a single input-output pair by employing a patch-based supervised learning strategy. In the context of sparse seismic inversion exceptional results were obtained following training based on a single training image. Interestingly, training with a simple synthetic example of relatively small size, consisting solely of horizontal lines and few faults \cite{Pereg:2019B} led to significantly improved results, which could be considered as zero-shot sparse seismic deconvolution. In the seismic exploration field, a similar framework was also used \cite{Biswas:2018,Pereg:2019B} to facilitate automatic migration velocity analysis, where up to $25\%$ of the acquired data is used for training and the system infers the rest of the missing velocities in the same survey. A similar framework has also been explored for optical coherence tomogeaphy (OCT) speckle suppression \cite{Pereg:2023B}, where one-shot learning models based on a recurrent neural net (RNN), as well as on a U-Net \cite{ronneberger:2015}, were both able to obtain state-of-the-art results. 

From a sparse representations \cite{Elad} point of view, a neural net of an encoder-decoder structure, maps its input into a latent space sparse code, that can then be used to extract the relevant information for the learning task, under the assumption that, with respect to a learned or known dictionary, \textit{everything that is not an event - is noise}. Patch-based dictionary learning \cite{Elad:2006,Papyan:2017} is a classic approach tackling image processing tasks. That said, previous works mostly focus on training the dictionary using patches from the corrupted image itself or training on a set of patches taken from a high-quality set of image training. 

This work investigates a supervised learning approach for image restoration tasks, given substantially limited access to ground truth training data. To this end, we employ a compact encoder-decoder framework, demonstrating the applicability and efficiency of one-shot learning for image deblurring and single-image super resolution. The proposed approach is potentially applicable to other learning architectures as well as other applications where the signal can be processed locally, such as speech and audio, video, medical imaging, natural language processing and more. Training efficiency of the proposed framework introduces a significant improvement. Namely, training takes less than 30 seconds on a GPU workstation, and few minutes on a CPU workstation, requiring minimal memory, thus significantly reducing the required computational resources. Inference time is also relatively low comparing with other image restoration recent works. In addition, to bring insight into a possible optimization mechanism of the proposed RNN-based patch-to-latent-space encoder, we observe that an RNN can be viewed as a sparse solver starting from an initial condition based on the previous time step. The
proposed interpretation presents a possible intuition for the mathematical functionality
enabling the use of RNNs for practical problem-solving. Finally, we provide a glimpse into a system mismatch case study with diverse Gaussian filters. 

To the best of our knowledge, this is the first research work validating the potential of one-shot image-to-image supervised learning framework for image restoration. Our work here is a substantial extension of our non-published work \cite{Pereg:2022}. 
 
\section{Preliminaries}\label{sec2}

\subsection{RNN Framework} \label{sec2.1}
Assume an observed sequence $\mathbf{y}=[\mathbf{y}_{0}, \mathbf{y}_{1}, ..., \mathbf{y}_{L-1}]^T$, $\mathbf{y}_t \in \mathbb{R}^{N \times 1}, \ t \in [0,{L-1}] $, and a corresponding output sequence $\mathbf{x}=[\mathbf{x}_{0}, \mathbf{x}_{1}, ..., \mathbf{x}_{L-1}]^T$, $\mathbf{x}_t \in \mathbb{R}^{P \times 1}$, where superscript
$^T$ denotes the transpose operation. The RNN forms a map $\mathcal{F} : \mathbf{y}\rightarrow \mathbf{z}$, from the input signal to the latent space variables, 
such that, for an input $\mathbf{y}_t$ and state $\mathbf{z}_t$ at time step $t$, the RNN output is generally formulated as 
$\mathbf{z}_t =  f(\mathbf{z}_{t-1},\mathbf{y}_t)$ \cite{Bengio:2013}.
Hereafter, we will focus on the specific parametrization:
\begin{equation}\label{3.2}
\mathbf{z}_t=\sigma(\mathbf{W}^T_{zy} \mathbf{y}_t + \mathbf{W}^T_{zz} \mathbf{z}_{t-1} + \textbf{b}),
\end{equation}
where $\sigma$ is an activation function, $\mathbf{W}_{zy} \in \mathbb{R}^{N \times n_{\mathrm{n}}}$ and $\mathbf{W}_{yy} \in \mathbb{R}^{n_{\mathrm{n}} \times n_{\mathrm{n}}}$ are weight matrices and $\mathbf{b}\in \mathbb{R}^{n_{\mathrm{n}} \times 1}$ is the bias vector, assuming $n_{\mathrm{n}}$ number of neurons in an RNN cell. At $t=0$ previous outputs are zero. Here, we use the ReLU activation function, $\mathrm{ReLU}(z)=\max\{0,z\}$. We then wrap the cell output, that is, the latent space vector $\mathbf{z}_t \in \mathbb{R}^{n_{\mathrm{n}} \times 1}$, with a fully connected layer such that the desired final output is $\mathbf{x}_t = \mathbf{W}^T_{xz}\mathbf{z}_t$ , where $\mathbf{W}_{xz} \in \mathbb{R}^{n_{\mathrm{n}} \times P}$.

RNNs are traditionally leveraged for processing of time-dependent signals, to predict future outcomes, and for natural language processing tasks such as handwriting recognition \cite{Graves:2009} and speech recognition \cite{Graves:2013}. In computer vision, RNNs are less popular, due to gradient exploding and gradient vanishing issues \cite{Bengio:2013}, and their expensive computational complexity compared with CNNs. The use of RNNs is less intuitive for computer vision tasks, because they are causal. Thus, Liang et al. (2015) \cite{Liang:2015} proposed recurrent convolutional networks (RCNNs) for object recognition. 
Pixel-RNN \cite{Van:2016} sequentially predicts pixels in an image along the two
spatial dimensions.

\subsection{Sparse Coding $\&$ Iterative Shrinkage Algorithms}
In sparse coding (SC), a signal $\mathbf{y}\in\mathbb{R}^{N \times 1}$ is modeled as a sparse superposition of feature vectors \cite{Elad, Chen:2001}. Formally, the observation signal obeys
$\mathbf{y}=\mathbf{D}\mathbf{z}$,
where $\mathbf{D}\in\mathbb{R}^{N \times M}$ is a dictionary of $M$ atoms $\mathbf{d}_i \in\mathbb{R}^{N \times 1}, \ i=1,...,M$, and $\mathbf{z} \in \mathbb{R}^{M \times 1}$ is a \textit{sparse} vector of the atoms weights. 
Over the years, many efforts have been invested in sparse coding, both in a noise free environment, or when allowing some error,
\begin{equation}\label{3.3}
\qquad \min_\mathbf{z} \|\mathbf{z}\|_1 \qquad 	\mathrm{s.t.}	 \qquad 	 \|\mathbf{y}-\mathbf{D}\mathbf{z}\|_2 \leq \varepsilon,
\end{equation}
where $\|\mathbf{z}\|_1 \triangleq \sum_i |z_i|$, $\|\mathbf{z}\|_2 \triangleq \sqrt{\sum_i z^2_i}$ and $\varepsilon$ is the residual noise or error energy.  
Further details on sparse coding are in the supplementary.

Consider the cost function, 
\begin{equation}\label{3.4}
f(\mathbf{z})= \frac{1}{2}\left\| \mathbf{y} - \mathbf{D} \mathbf{z} \right\|_2^2 + \lambda {\left\| {\mathbf{z}}\right\|_1},
\end{equation}
for some scalar $\lambda>0$. Following Majorization Minimization (MM) strategy, we can build a surrogate function \cite{Daubechies:2004,Elad}
\begin{equation}\label{3.5}
Q(\mathbf{z},\mathbf{z}_\theta) = f(\mathbf{z}) + d(\mathbf{z},\mathbf{z}_\theta) = 
\frac{1}{2}\left\| \mathbf{y} - \mathbf{D} \mathbf{z} \right\|_2^2 
+
 \lambda {\left\| {\mathbf{z}}\right\|_1} +
\frac{c}{2} \|\mathbf{z}-\mathbf{z}_\theta \|_2^2-\frac{1}{2} \| \mathbf{D} \mathbf{z} -   \mathbf{D} \mathbf{z}_\theta\|_2^2.
\end{equation}
The parameter $c$ is chosen such that the added expression 
\begin{equation}\label{3.6}
d(\mathbf{z},\mathbf{z}_\theta)=Q(\mathbf{z},\mathbf{z}_\theta)-f(\mathbf{z})= \frac{c}{2} \|\mathbf{z}-\mathbf{z}_\theta \|_2^2-\frac{1}{2} \| \mathbf{D} \mathbf{z} -   \mathbf{D} \mathbf{z}_\theta\|_2^2
\end{equation}
is strictly convex, requiring its Hessian to be positive definite, $c \mathbf{I} - \mathbf{D}^T\mathbf{D} \succ \mathbf{0}$. Therefore $ c > \| \mathbf{D}^T\mathbf{D} \|_2 = \alpha_{max}(\mathbf{D}^T\mathbf{D})$, i.e., greater than the largest eigenvalue of the coherence matrix $\mathbf{D}^T\mathbf{D}$. The term $d(\mathbf{z},\mathbf{z}_\theta)$ is a measure of proximity to a previous solution $\mathbf{z}_\theta$. If the vector difference $\mathbf{z}-\mathbf{z}_\theta$ is spanned by $\mathbf{D}$, the distance drops to nearly zero. Alternatively, if $\mathbf{D}$ is not full rank and the change $\mathbf{z}-\mathbf{z}_\theta$ is close to the null space of $\mathbf{D}$, the distance is simply the approximate Euclidean distance between the current solution to the previous one.
The sequence of iterative solutions minimizing $Q(\mathbf{z},\mathbf{z}_\theta)$ instead of $f(\mathbf{z})$, is generated by the recurrent formula
$\mathbf{z}_{\theta+1} =  \arg \min_{\mathbf{z}} Q(\mathbf{z},\mathbf{z}_\theta)$,
where $\theta \in \mathbb{N}$ is the iteration index.
We can find a closed-form solution for its global minimizer
that can be intuitively viewed as an iterative projection of the dictionary on the residual term, starting from the initial solution that is a thresholded projection of the dictionary on the observation signal, assuming $\mathbf{z}_0=\mathbf{0}$:
\begin{equation}\label{3.7}
\mathbf{z}_{\theta+1} =  \mathcal{S}_{\frac{\lambda}{c}}\Big(
\frac{1}{c} \mathbf{D}^T
(\mathbf{y}-\mathbf{D}\mathbf{z}_\theta)
+ \mathbf{z}_\theta \Big)=\mathcal{S}_{\frac{\lambda}{c}}
\Big(\frac{1}{c}\mathbf{D}^T\mathbf{y}+
\big(\mathbf{I}-\frac{1}{c} \mathbf{D}^T\mathbf{D}\big) \mathbf{z}_\theta\Big),
\end{equation}
which could be intuitively comprehended as
\begin{equation}\label{2.11}
\mathbf{z}_{\theta+1} =  \mathcal{S}_{\frac{\lambda}{c}}\Big(
\frac{1}{c} 
\overbrace{\mathbf{D}^T}^\text{project on dictionary}
\underbrace{(\mathbf{y}-\mathbf{D}\mathbf{z}_\theta)}_\text{residual term} \quad
 + \underbrace{\mathbf{z}_\theta}_\text{add to current solution} \Big).
\end{equation}
where the $\mathcal{S}_\beta(z)=(|z|-\beta)_{+} \mathrm{sgn}(z)$ is the soft threshold operator.
It is guaranteed that the cost is decreasing with each iteration.  
Assuming the constant $c$ is large enough, it was shown in \cite{Daubechies:2004}, that (\ref{3.7}) is guaranteed to converge to its global minimum.
This approach can also be viewed as a proximal-point algorithm \cite{Combettes:2005}, or as a simple projected gradient descent algorithm. 
Over time, faster extensions have been suggested, such as: Fast-ISTA (FISTA) \cite{Beck:2009}, and Learned-ISTA (LISTA) \cite{Lecun:2010}, Ada-LISTA \cite{Aberdam:2022} and RFN-ITA \cite{Pereg:2021}. 
LISTA is formulated as:
\begin{equation}\label{3.8}
\mathbf{z}_{\theta+1} =  
\mathcal{S}_{\frac{\lambda}{c}}
\big(\mathbf{W}\mathbf{y}+
\mathbf{S}\mathbf{z}_\theta\big).
\end{equation}
$\mathbf{W}$ and $\mathbf{S}$ are learned over a set of training samples
$\{\mathbf{y}_i,\mathbf{z}_i\}_{i=1}^m$. Note that $\mathbf{W}$ and $\mathbf{S}$ re-parametrize the matrices $\frac{1}{c}\mathbf{D}^T$ and $\big(\mathbf{I}-\frac{1}{c}\mathbf{D}^T\mathbf{D}\big)$, respectively. 

\section{RNN Analyzed via Sparse Coding}\label{sec4}
Observing the similar structure of (\ref{3.2}) and (\ref{3.8}), we redefine the cost function (\ref{3.4})
\begin{equation}\label{4.1}
f(\mathbf{z}_t)= \frac{1}{2}\left\| \mathbf{y}_t - \mathbf{D} \mathbf{z}_t \right\|_2^2 + \lambda {\left\| {\mathbf{z}_t}\right\|_1}.
\end{equation}
Now, building a surrogate function
\begin{flalign}\label{4.2}
\nonumber 
Q(\mathbf{z}_t,\mathbf{z}_{t-1}) & = f(\mathbf{z}_t) + d(\mathbf{z}_t,\mathbf{z}_{t-1})  \\
& = \frac{1}{2}\left\| \mathbf{y}_t - \mathbf{D} \mathbf{z}_t \right\|_2^2 
+
 \lambda {\left\| {\mathbf{z}_t}\right\|_1} +
\frac{c}{2} \|\mathbf{z}_t-\mathbf{z}_{t-1} \|_2^2-\frac{1}{2} \| \mathbf{D} \mathbf{z}_t -   \mathbf{D} \mathbf{z}_{t-1}\|_2^2,
\end{flalign}
where the added term $d(\mathbf{z}_t,\mathbf{z}_{t-1})$
represents the distance between the \textit{current solution} and the \textit{previous solution} at the \textit{preceding time step} (rather than the previous iteration). We now have,
\begin{equation}\label{4.4}
\mathbf{z}_{t} =  \mathcal{S}_{\frac{\lambda}{c}}\Big(
\frac{1}{c} \mathbf{D}^T
(\mathbf{y}_t-\mathbf{D}\mathbf{z}_{t-1})
+ \mathbf{z}_{t-1}\Big),
\end{equation}
which in its learned version can be re-parametrized as,
\begin{equation}\label{4.5}
\mathbf{z}_{t} =  \mathcal{S}_{\beta} \Big(\mathbf{W}^T_{zy}\mathbf{y}_t+\mathbf{W}^T_{zz}\mathbf{z}_{t-1}\Big).
\end{equation} 
Clearly, (\ref{4.5}) is equivalent to (\ref{3.2}). In other words, \textit{a RNN can be viewed as an unfolding of one iteration of a learned sparse coder, based on an assumption that the solution at time $t$ is close to the solution in time $t-1$}.
In subsection \ref{sec2.1}, the RNN state encodes the vector to a latent space.
Then the linear projection (fully connected network) decodes the latent variable back to a space of the 
required dimensions.
Given this interpretation, it may be claimed that RNN's use should not be restricted to data with obvious time or depth relations. The RNN merely serves as an encoder providing a rough estimation of the sparse code of the input data. The RNN's memory serves the optimization process by starting the computation from a closer solution. Thus, placing the initial solution in a ``close neighborhood" or close proximity, and helping the optimization gravitate more easily towards the latent space sparse approximation. Clearly, convergence is not guaranteed. 

\section{Patch-based Learning}\label{sec4}

\textbf{Patch-based Learning via RNN.}
The preliminary setting described in this subsection was previously employed for several applications in seismic imaging and medical imagining \cite{Biswas:2018,Pereg:2019B,Pereg:2020,Pereg:2023B}.
The proposed patch-based framework is not restricted to RNNs, and can be replaced with a different patch-based architecture, such as a U-Net \cite{ronneberger:2015}.
The description below is formulated for two-dimensional (2D) input signals, but can be easily adapted to other input data dimensions. We use similar definitions and notations as previously described in \cite{Pereg:2019B,Pereg:2020}.

\textbf{Definition 1} (Analysis Patch): \label{Def1} 
We define an \textit{analysis patch} as a 2D patch of size $L_\mathrm{t}\times N_\mathrm{x}$ enclosing $L_\mathrm{t}$ time (rows) samples of $N_\mathrm{x}$ consecutive neighboring columns of the observed image $\mathbf{Y}\in\mathbb{R}^{L_\mathrm{r} \times J}$. Assume $\{n_\mathrm{L},n_\mathrm{R} \in \mathbb{N} :n_\mathrm{L}+n_\mathrm{R}=N_\mathrm{x}-1\}$. The analysis patch $\mathbf{A}^{(i,j)}$ associated with an image point at location $(i,j)$, such that element $(k,l)$ of $\mathbf{A}^{(i,j)}$ is
\begin{equation*}
\mathbf{A}^{(i,j)}_{k,l} = \big\{ Y[i-k,j+l] \ : \  k,l \in \mathbb{Z}, \ 0 \leq k \leq L_t-1 , \ -n_L \leq l \leq n_R \big\} .
\end{equation*}
The analysis patch moves through the input image and produces the expected output image point-values $\mathbf{X}\in\mathbb{R}^{L_\mathrm{r} \times J}$.

\textbf{Patch2Pixel RNN.} Each analysis patch is mapped to an output segment ($P=1$), and the last point in each segment is chosen as the predicted output pixel. In other words, the model's input is an analysis patch $\mathbf{y} = \mathbf{A}^{(i,j)}\in\mathbb{R}^{L_\mathrm{t} \times N_\mathrm{x}}$, and the model's output is the pixel $X[i,j]$ associated with the corresponding analysis patch. 
Namely, a time-step input is a vector of $N_\mathrm{x}$ neighboring pixels of the corresponding time (depth). Thus, in this setting 
$N=N_\mathrm{x}$ and
$\mathbf{y}_t = \big[ Y[t,j-n_L], ...,Y[t,j+n_R]\big] ^T$.
The size of the output vector $\mathbf{z}_t$ is one ($P=1$), such that $\mathbf{x}$ is the corresponding output image segment, 
$\mathbf{x} = \big[ X[i-(L_t-1),j], ... , X[i,j] \big] ^T$.
Lastly, the first $L_t-1$ output values of $\mathbf{x}$ are discarded, while the last pixel, $\mathbf{x}_{L_t}$ is kept as the predicted pixel $\hat{X}[i,j]$. 
As the analysis patch scans the image, all predicted pixels are obtained in a similar manner.

\begin{figure}
\centering
\includegraphics[width=13.75cm,height=8.2cm]{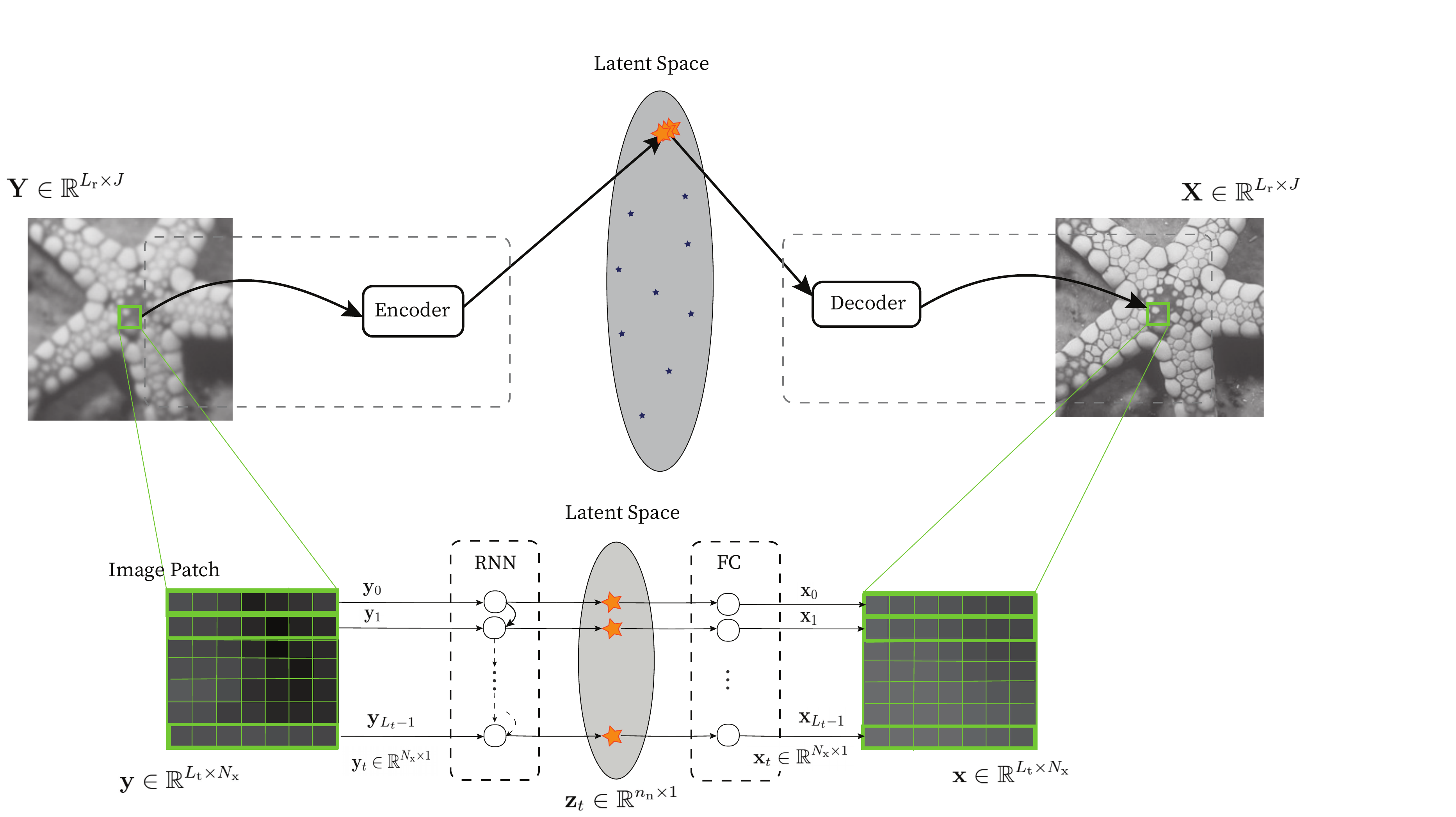}
\caption{Illustration of the proposed patch-to-patch RNN encoder-decoder.}
\label{rnn_fig}
\vspace{-0.75cm}
\end{figure}

\textbf{Patch2Patch RNN.} 
Given a degraded image $\mathbf{Y}$, and an original image $\mathbf{X}$, an alternative approach is to process overlapping patches, restore
each patch separately and ﬁnally average the obtained patches back into an  image. 
Averaging of overlapping patch-estimates is a common approach in patch-based algorithms \cite{lebrun:2012,lebrun:2013}, such as expected patch log-likelihood (EPLL) \cite{hurault:2018}. As we are averaging over a set of diﬀerent estimates, it is also expected to increase SNR. In this setting, the input analysis patch remains  $\mathbf{y} = \mathbf{A}^{(i,j)} \in\mathbb{R}^{L_\mathrm{t} \times N_\mathrm{x}}$. While, the output in no longer a 1D segment, but a corresponding output 2D patch, $N=N_\mathrm{x}$, 
$\mathbf{x}_t = \big[X[t,j-n_L], ...,X[t,j+n_R]\big]^T$ such that $P=N_\mathrm{x}$.
The RNN-based patch2patch setting is illustrated in Fig.~\ref{rnn_fig}.

It has been observed that over-parametrized neural networks generalize better \cite{Lecun:2018} and that dictionary recovery may be facilitated with over-realized models \cite{Sulam:2022}. 
In accordance with these observations, we note that here, the number of RNN hidden units is significantly larger than the dimension of the input, $n_n >> N_\mathrm{x}$. Namely, the latent space dimension is 3 orders of magnitude larger than the input dimension. In practice, the encoder's latent space activations are sparse (without additional sparse constraints \cite{NG:2011,Vincent:2008}). 

An obvious downside of the use of a one-directional RNN  is the underlying assumption that the ``time" relation is causal in the vertical direction of the image, which might not be the case for some signals, yet has shown to be effective in natural images and tomographic applications. A different configuration that might serve as a partial remedy is possible. For example, one can define each time input $\mathbf{x}_t$ as a patch (rather than a row-segment in a patch), or move through the image in a different direction. 

\textbf{Patch2Patch U-Net.} 
The U-Net is a convolutional neural network that was developed for biomedical image segmentation \cite{ronneberger:2015} and obtained state of the art results in numerous applications. Since the U-Net's input size is flexible,  
we can employ a U-Net patch-based one-shot learning framework by training with random patches cropped from a \textit{single} input-output pair. The U-Net is then applied to an image of the user's desired size.

\textbf{Incremental Generative Adversarial Network.}
Image restoration algorithms are normally evaluated by distortion measures (e.g. PSNR, SSIM) or by perceptual quality scores, 
which are at odds with each other \cite{Blau:2018}. 
Consequently, several previous works adopt a two-stages training \cite{Ledig:2017,Pereg:2023B}. The first stage trains the generator to optimize a content loss. While, in the second stage, initialized by the generator’s pre-trained weights, we train both a generator $G$ and a discriminator $D$. 
Therefore, we add a second stage of training with a combined MSE and adversarial loss,
$\mathcal{L}_\mathrm{G}  = \mathcal{L}_\mathrm{MSE} + \lambda \mathcal{L}_\mathrm{ADV}$,
where $\lambda$ is a constant balancing the losses. 
To this end, we employ a patch-based discriminator, that consists simply of 2 fully-connected layers. 
The generator $G$ can be either a patch-to-patch RNN-based predictor or a U-Net trained with patches from a single image. We will refer to these methods as RNN-GAN and UNET-GAN, respectively.

\section{Numerical Experiments}\label{sec5}

We study the proposed one-shot image-to-image framework for three implementations that we will refer to as: RNN, RNN-GAN and UNET-GAN.
We compare our results with PnP-PGD and RED-SD, applied with the DnCNN denoiser \cite{Cohen:2021,Zhang:2017}.

\textbf{Datasets $\&$ Training.}
For training the UNET-GAN and the RNN-GAN we used Lena color image of size $512\times 512$, and an analysis patch size of $16 \times 16$ and $15 \times 15$ respectively. For the RNN we used  Peppers color image of size $512 \times 512$, and an analysis patch size of $9 \times 9$ for image deblurring, and $11 \times 11 $ for super resolution. Generally speaking the RNN is relatively sensitive to the choice of the training image.
The minimal analysis patch size recommended is $7 \times 7$.  
The minimal training image size we experimented with is $128 \times 128$.
For the UNET-GAN, the generator is a standard U-Net that consists of four contracting layers with unit stride ($s=1$) to avoid the known cross-hatch artifacts normally occurring due to larger strides \cite{Deng:2022}. 
For the RNN-GAN the generator is the RNN-based patch2patch encoder-decoder model described above.
For all experiments, we set the RNN number of neurons as $n_\mathrm{n} = 1000$. Increasing the
number of neurons did not improve the results signiﬁcantly, but increases training time. Notably, the U-Net has about $8.2 \times 10^6$ parameters, that is $\sim8$ times the RNN number of parameters, and $\sim16$ times the DnCNN number of parameters. Note that all proposed architectures are trained with a single example pair of degraded image and its corresponding ground truth. 
We trained the generator with $\ell_1$ loss, typically applied for deblurring and SR tasks \cite{Shocher:2018}, for 25 epochs. Then we train the generator and the discriminator for additional 80 epochs. The RNN is trained for 45 epochs only. 
Note that this training fits the training image almost perfectly. It is possible to train with substantially less epochs without significant degradation in the model's performance. The discriminator consists of only 2 fully-connected layers.
We used the Adam-optimizer \cite{Kingma:2014} with $\beta_1=0.5$, $\beta_2=0.9$. The initial learning rate is $10^{-4}$. 
Training was executed using a Laptop GPU NVIDIA RTX 3500 equipped with only 12 GBs of video memory. 
Training the RNN and RNN-GAN requires substantially less GPU-memory (less than $0.4$GB) and can be performed using smaller laptop GPUs as indicated in Table~\ref{Table 3}.
All frameworks may also be trained using solely a CPU-based computational resource. 
For the task of patch2patch super resolution we employ residual learning for the RNN-GAN and the UNET-GAN. 
We used two known datatsets for testing, namely: 
Set11 provided by the authors of \cite{romano:2017}, which consists of 11 common color images
(butterfly, starfish, etc), and 500 natural images from the public Berkeley segmentation dataset (BSD500) \cite{Martin:2001}. 
All pixels values in the set are in the range of [0,255].

\textbf{Inverse Problems.}
For the task of image deblurring, the images were convolved with a 2D Gaussian PSF of size $25 \times 25$ with standard deviation of  $\sigma=1.6$, with additional white Gaussian noise (WGN) of level $\sigma_n=\sqrt{2}$. For the task of super resolution the image is convolved with the same 2D Gaussian filter, scaled down by a factor of 3 in both axes, and contaminated with WGN of $\sigma_n=\sqrt{2}$. 
Similarly to \cite{romano:2017}, RGB images are converted to YCbCr color-space, inversion is applied to the luminance channel, and the result is converted back to the RGB domain. As a figure of merit, we used the peak signal to noise ration (PSNR) and the SSIM, both computed on the estimated luminance channel of the ground truth and the estimated image. 

\begin{figure*}[t!]
    \begin{subfigure}[t]{0.16\textwidth}
        \includegraphics[width=0.95\linewidth]{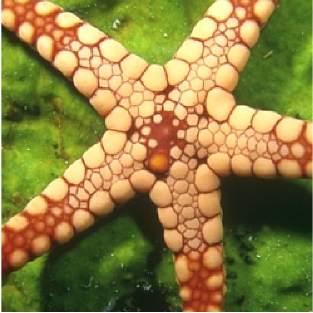}
				\caption{}
    \end{subfigure}%
		\hfill 
		\begin{subfigure}[t]{0.16\textwidth}
        \includegraphics[width=0.95\linewidth]{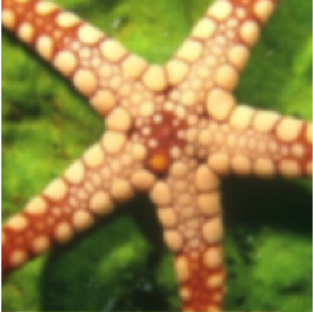}
				\caption{}
    \end{subfigure}%
		\hfill 
    \begin{subfigure}[t]{0.16\textwidth}
        \includegraphics[width=0.95\linewidth]{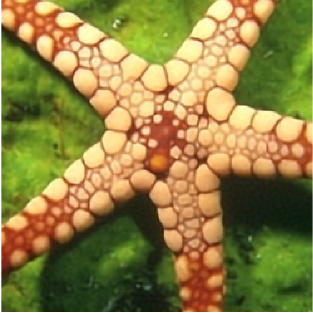}
				\caption{}
    \end{subfigure}%
		\hfill 
		\begin{subfigure}[t]{0.16\textwidth}
        \includegraphics[width=0.95\linewidth]{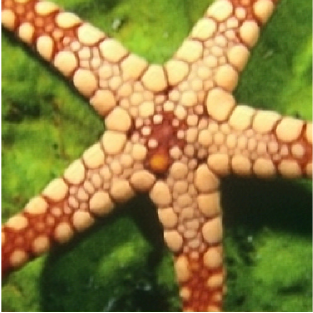}
				\caption{}
    \end{subfigure}%
		\hfill
		\begin{subfigure}[t]{0.16\textwidth}
				\includegraphics[width=0.95\linewidth]{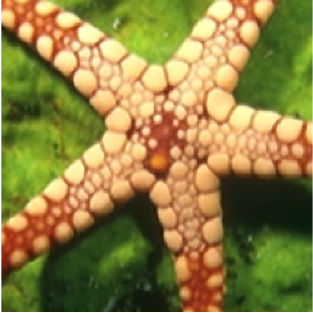}
				\caption{}
		\end{subfigure}%
		\hfill
		\begin{subfigure}[t]{0.16\textwidth}
				\includegraphics[width=0.95\linewidth]{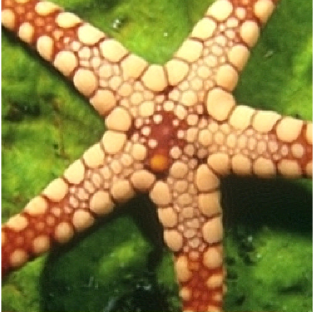}
				\caption{}
    \end{subfigure}%
\caption{{\small Visual comparison of deblurring of the image starfish: (a) Ground truth; (b) input, 24.8dB; (c) RED-SD, 32.42dB;
(d) RNN, 29.18dB; (e) RNN-GAN, 29.11dB; (f) UNET-GAN, 30.94 dB.}}
\label{fig6} 
\end{figure*}

\begin{figure*}[t!]
    \begin{subfigure}[t]{0.16\textwidth}
        \includegraphics[width=0.95\linewidth]{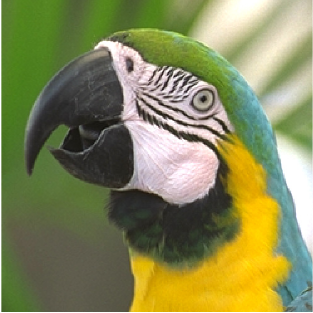}
				\caption{}
    \end{subfigure}%
		\hfill 
		\begin{subfigure}[t]{0.16\textwidth}
        \includegraphics[width=0.95\linewidth]{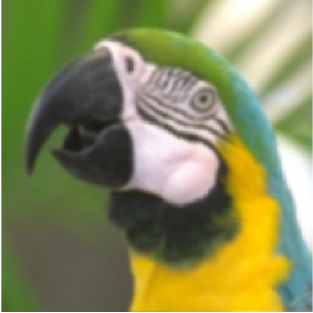}
				\caption{}
    \end{subfigure}%
		\hfill 
    \begin{subfigure}[t]{0.16\textwidth}
        \includegraphics[width=0.95\linewidth]{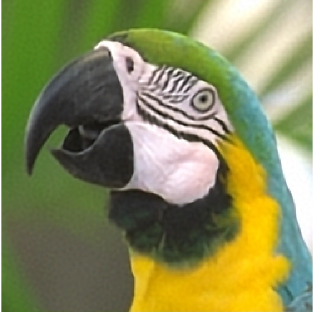}
				\caption{}
    \end{subfigure}%
		\hfill 
		\begin{subfigure}[t]{0.16\textwidth}
        \includegraphics[width=0.95\linewidth]{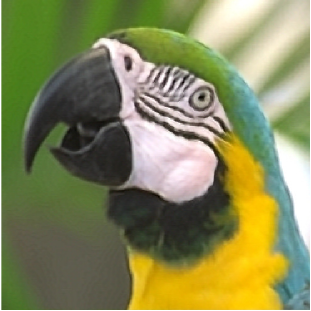}
				\caption{}
    \end{subfigure}%
		\hfill
		\begin{subfigure}[t]{0.16\textwidth}
				\includegraphics[width=0.95\linewidth]{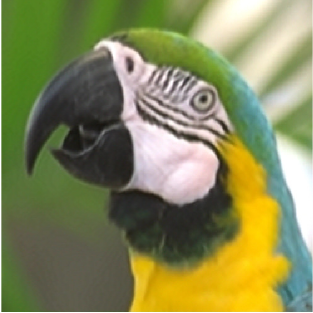}
				\caption{}
    \end{subfigure}%
		\hfill
		\begin{subfigure}[t]{0.16\textwidth}
				\includegraphics[width=0.95\linewidth]{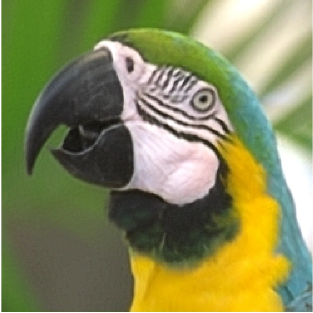}
				\caption{}
    \end{subfigure}%
\caption{{\small Visual comparison of deblurring of the parrot image: (a) Ground truth; (b) input, 25.33dB; (c) RED-SD, 33.18dB; (d) RNN, 30.98dB; (e) RNN-GAN, 30.30dB; (f) UNET-GAN, 32.22dB.}}
\label{fig7} 
\end{figure*}

\begin{figure*}[h]
    \begin{subfigure}[t]{0.18\textwidth}
        \includegraphics[width=0.95\linewidth]{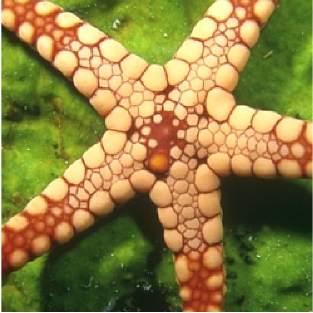}
				\caption{}
    \end{subfigure}%
		\hfill 
		\begin{subfigure}[t]{0.18\textwidth}
        \includegraphics[width=0.95\linewidth]{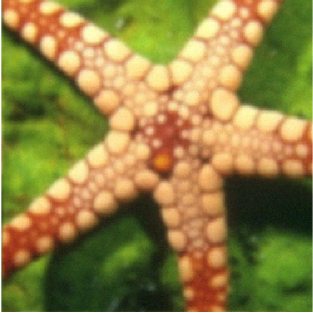}
				\caption{}
    \end{subfigure}%
		\hfill 
    \begin{subfigure}[t]{0.18\textwidth}
        \includegraphics[width=0.95\linewidth]{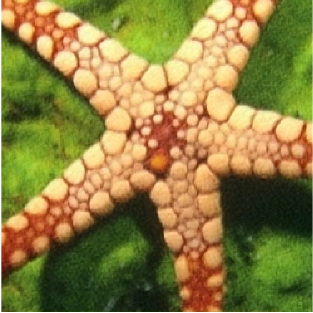}
				\caption{}
    \end{subfigure}%
		\hfill 
		\begin{subfigure}[t]{0.18\textwidth}
        \includegraphics[width=0.95\linewidth]{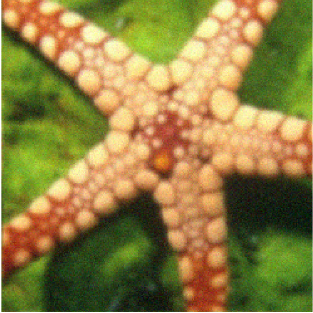}
				\caption{}
		\end{subfigure}%
		\hfill 		
		\begin{subfigure}[t]{0.18\textwidth}
        \includegraphics[width=0.95\linewidth]{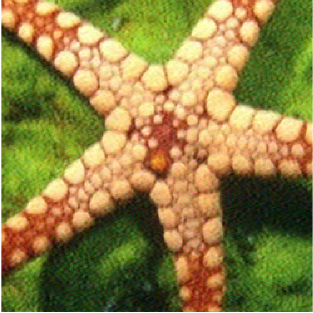}
				\caption{}
    \end{subfigure}%
\caption{{\small Visual comparison of deblurring of the image starfish for different noise levels: (a) Ground truth; (b) input, 24.25dB, $\sigma_n=4.24$; (c) RNN, 28.07dB ;
(d) input, 23.02dB, $\sigma_n=9.90$; (e) RNN-GAN, 26.36dB .}}
\label{figD2} 
\end{figure*}

\begin{figure*}[t!]
    \begin{subfigure}[t]{0.16\textwidth}
        \includegraphics[width=0.95\linewidth]{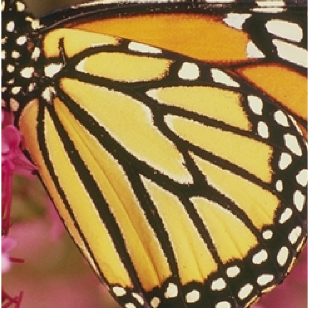}
				\caption{}
    \end{subfigure}%
		\hfill 
		\begin{subfigure}[t]{0.16\textwidth}
        \includegraphics[width=0.95\linewidth]{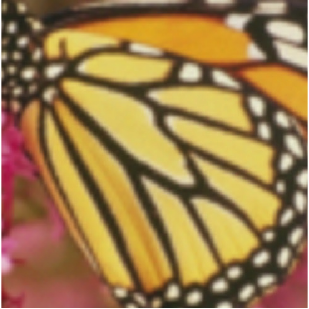}
				\caption{}
    \end{subfigure}%
		\hfill 
    \begin{subfigure}[t]{0.16\textwidth}
        \includegraphics[width=0.95\linewidth]{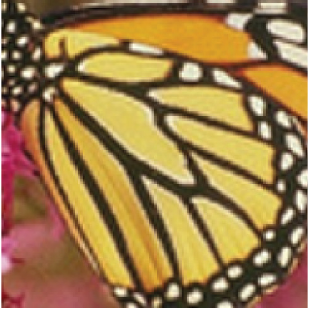}
				\caption{}
    \end{subfigure}%
		\hfill 
		\begin{subfigure}[t]{0.16\textwidth}
        \includegraphics[width=0.95\linewidth]{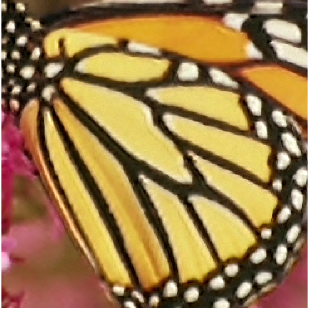}
				\caption{}
    \end{subfigure}%
		\hfill
		\begin{subfigure}[t]{0.16\textwidth}
				\includegraphics[width=0.95\linewidth]{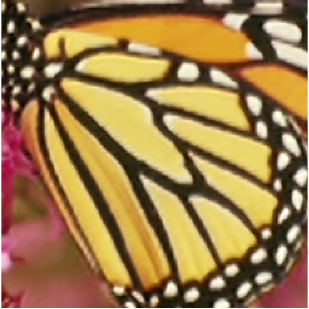}
				\caption{}
    \end{subfigure}%
		\hfill
		\begin{subfigure}[t]{0.16\textwidth}
				\includegraphics[width=0.95\linewidth]{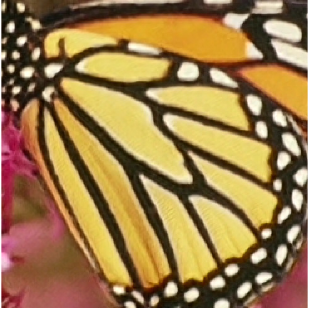}
				\caption{}
    \end{subfigure}%
\caption{{\small Visual comparison of super resolution of the image butterfly: (a) Ground truth; (b) Bicubic interpolation, 19.44dB; (c) RED-SD (DnCNN), 23.57dB; (d) RNN, 22.84dB; (e) RNN-GAN, 22.45dB; (f) UNET-GAN, 22.83dB.}}
\label{fig_sr} 
\end{figure*}

\begin{figure*}[t!]
    \begin{subfigure}[t]{0.166\textwidth}
        \includegraphics[width=0.95\linewidth]{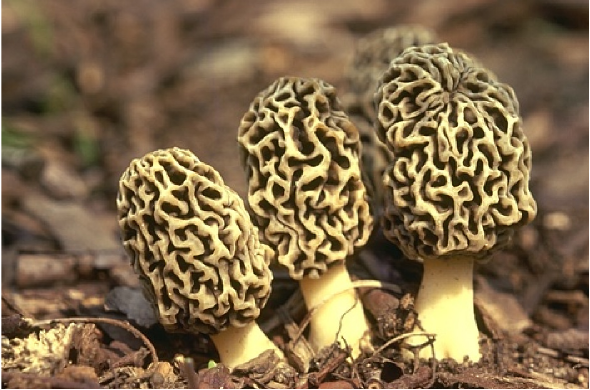}
				\caption{}
    \end{subfigure}%
		\hfill 
		\begin{subfigure}[t]{0.166\textwidth}
        \includegraphics[width=0.95\linewidth]{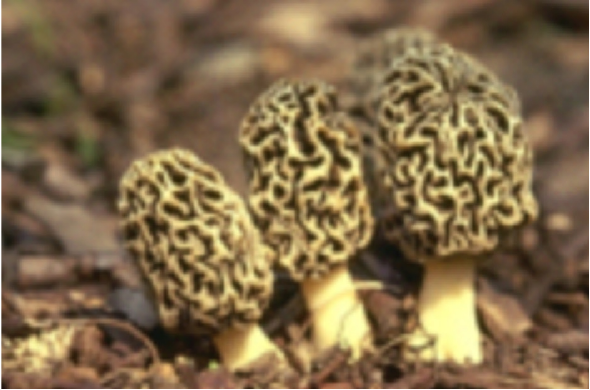}
				\caption{}
    \end{subfigure}%
		\hfill 
    \begin{subfigure}[t]{0.166\textwidth}
        \includegraphics[width=0.95\linewidth]{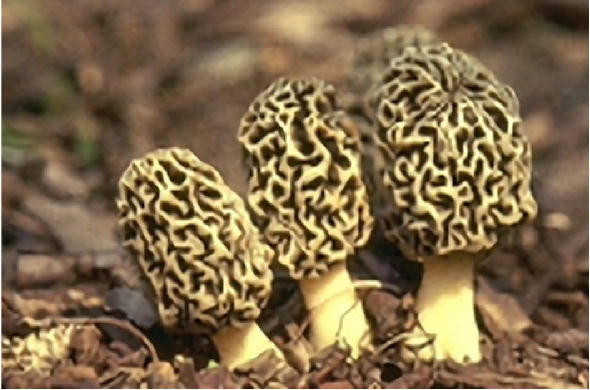}
				\caption{}
    \end{subfigure}%
		\hfill 
		\begin{subfigure}[t]{0.166\textwidth}
        \includegraphics[width=0.95\linewidth]{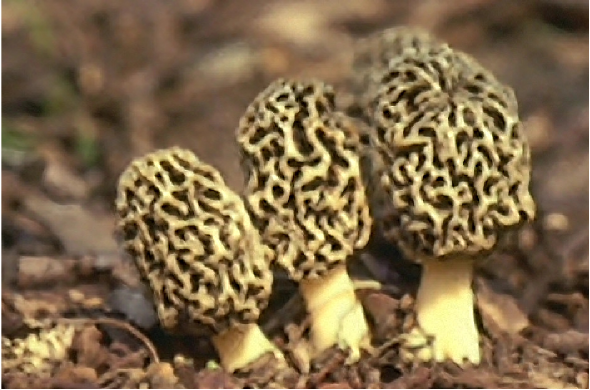}
				\caption{}
    \end{subfigure}%
		\hfill
		\begin{subfigure}[t]{0.166\textwidth}
				\includegraphics[width=0.95\linewidth]{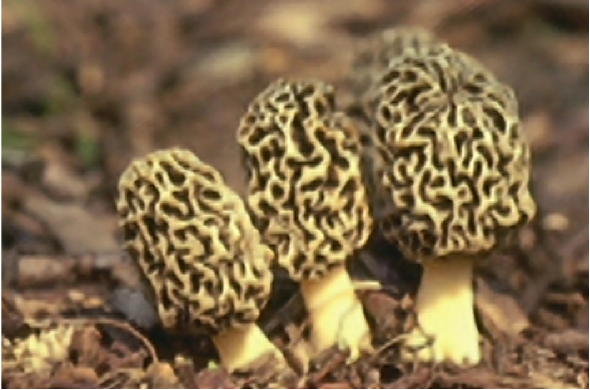}
				\caption{}
    \end{subfigure}%
		\hfill
		\begin{subfigure}[t]{0.166\textwidth}
				\includegraphics[width=0.95\linewidth]{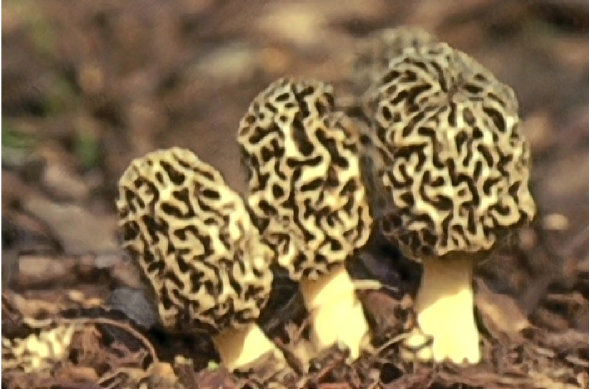}
				\caption{}
    \end{subfigure}%
\caption{{\small Visual comparison of super resolution of example image from BSD500: (a) Ground truth; (b) Bicubic interpolation, 19.95dB; (c) RED-SD, 26.19dB; 
(d) RNN 24.86dB; (e) RNN-GAN, 23.22dB; (f) UNET-GAN, 23.39dB.}}
\label{fig_sr_91} 
\end{figure*}

Tables~\ref{Table A}-\ref{Table B} present the average PSNR and SSIM scores obtained, compared with state-of-the-art image restoration methods. Note that most deblurring methods require prior knowledge of the degradation process. Whereas the proposed approach requires only one example of the degraded image and its corresponding ground truth. For the RNN, the best scores were obtained by training with either one of the images: butterfly, boats, parrot, starfish and peppers. The RNN-GAN and UNET-GAN scores were uniform across all training images. Generally speaking, among the one-shot learning frameworks, UNET-GAN yields the best scores, but its training takes longer and requires more parameters. RNN yields slightly higher PSNR scores comparing with RNN-GAN, but in some cases RNN-GAN obtains slightly higher SSIM scores. We did not observe significant differences between the two methods under this noise level. 
Figures~\ref{fig6}-\ref{fig_sr_91} present visual examples of the proposed methods.
We observe that, contrary to common belief, our one-shot learning framework produces results that are comparable with 
other state-of-the-art image restoration methods, despite being trained only with a single input-output pair.

\begin{table*}[h!]
\begin{center}
\caption{Average PSNR / SSIM obtained for Gaussian Deblurring}
\label{Table A}
{\footnotesize

  \begin{tabular}{ |c|c|c|ccc|}
    \hline
		 Datasets 		&PnP-PGD  		&RED-SD  					& \multicolumn{3}{c|}{One-Shot learning} \\  
		 						  &DnCNN 				&DnCNN						&RNN     		&RNN-GAN   			&UNET-GAN

			\\ \hline \hline

		Set11      &28.81 / 0.86		&28.69 / 0.82		 & 28.59 / 0.86			& 27.81 /	0.85		& 29.12 / 0.88 	\\ \hline
		
		BSD500     &28.79 / 0.84		&28.32 / 0.79    & 28.38 / 0.83			&	27.52 / 0.82		& 28.42 / 0.84  \\ \hline

  \end{tabular} 
} \\
\end{center}
\vspace{-0.2cm}
\end{table*}

\begin{table*}[h!]
\begin{center}
\caption{Average PSNR / SSIM obtained for Super Resolution}
\label{Table B}
{\footnotesize

  \begin{tabular}{ |c|c|c|ccc|}
    \hline
		 Datasets 		&PnP-PGD  		&RED-SD  					& \multicolumn{3}{c|}{One-Shot learning} \\  
		 						  &DnCNN 				&DnCNN						&RNN     		&RNN-GAN   			&UNET-GAN 		
		
			\\ \hline \hline
				
		 Set11     & 26.20 / 0.77			&	26.14 / 0.76			& 25.77 /	0.75		& 25.79 / 0.76 		& 25.69 / 0.76   \\ \hline
		
		 BSD500 	 & 26.36 / 0.75			&	26.24 / 0.73			&	26.23 / 0.71		& 25.78 / 0.71		& 25.58 / 0.73   	\\ \hline
									 
  \end{tabular} 
} \\
\end{center}
\end{table*}

\begin{figure}[h!]
\centering
\includegraphics[width=4.5cm, height=3.5cm]{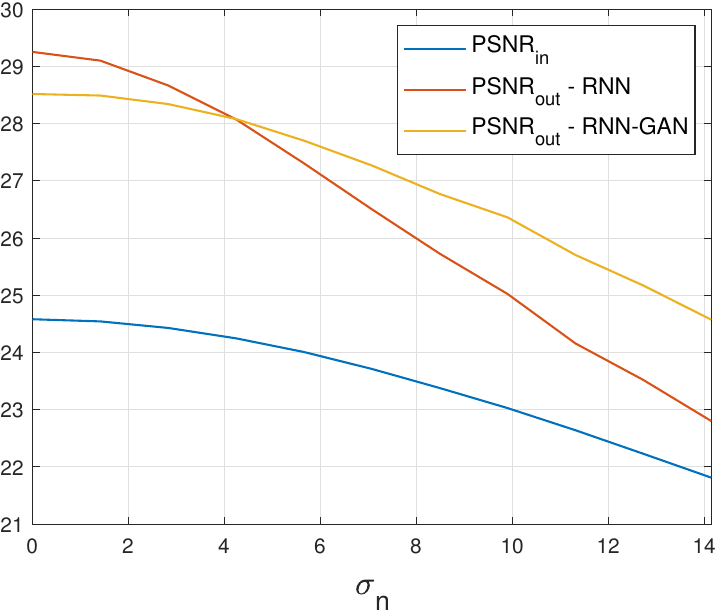} \\
\caption{{\small PSNR [dB] for image starfish vs input noise variance}.}
\label{figD1}
\vspace{-10pt}
\end{figure}

To test the systems' performance under more challenging setting we tested the RNN and RNN-GAN with increasing additive noise variance $\sigma_n$ in the range $[0,10\sqrt{2}]$. We trained the network with a \textit{single} example of the blurred image boat contaminated by Gaussian noise with variance $\sigma_n=\sqrt{2}$. Figure~\ref{figD1} show the evolution of the PSNR scores for RNN and RNN-GAN with increasing additive noise variance $\sigma_n \in [0,10\sqrt{2}]$. As the noise level increases, the RNN-GAN suppresses noise better. Figure~\ref{figD2} provides additional examples of the reconstruction of image starfish for $\sigma_n=2\sqrt{2}$ and $\sigma_n=7\sqrt{2}$. 

\begin{figure}[h!]
\centering
\includegraphics[width=4.5cm, height=3.25cm]{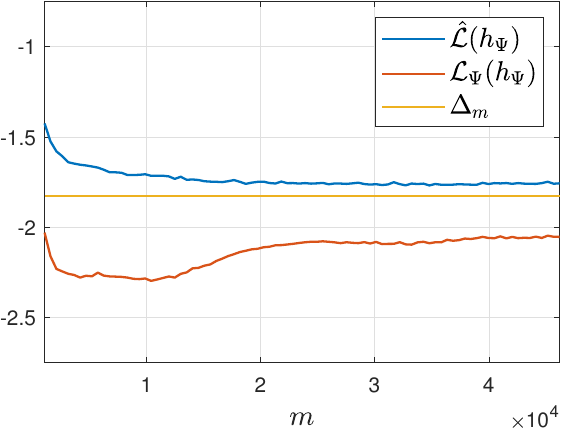}\\
\caption{ Recovery error $\hat{\mathcal{L}}(h_{\Psi})$ and training error $\mathcal{L}_{\Psi}(h_{\Psi}) \leq \Delta_m$ ($\log_{10}$ scale) as a function of the sample size $m$.}
\label{figD4}
\end{figure}

Figure~\ref{figD4} shows the evolution of the recovery error $\hat{\mathcal{L}}(h_{\Psi})$, with increasing sample size $m$, i.e., number of patches used during training,
where $h_{\Psi}$ denotes the trained predictor. 
The RNN network was trained with varying sample sizes while stopping the training at empirical risk (error) $\mathcal{L}_{\Psi}(h_{\Psi}) \leq \Delta_m$, where
\begin{equation*}
\mathcal{L}_{\Psi}(h_{\Psi}) = \frac{1}{m} \sum_{i=1}^m \ell(h_{\Psi}(\mathbf{y}_i),\mathbf{x}_i),
\end{equation*}
is computed over a training set $\Psi=\{\mathbf{y}_{i},\mathbf{x}_{i}\}_{i=1}^m$ of input-output paired patches, with squared $\ell_2$ loss.
The measured recovery error is computed over the remaining 9 non-training images, in the dataset used in \cite{Cohen:2021}, as an empirical approximation for the generalization error. The training data consists of patches belonging to a single image (starfish). As can be seen, $\hat{\mathcal{L}}(h_{\Psi})$ converges for a relatively small number of patches. Thus training can achieve reasonable generalization with patches extracted from an image with size as small as $128 \times 128$.

\begin{figure*}[t!]

    \begin{subfigure}[t]{0.2\textwidth}
        \includegraphics[width=0.95\linewidth]{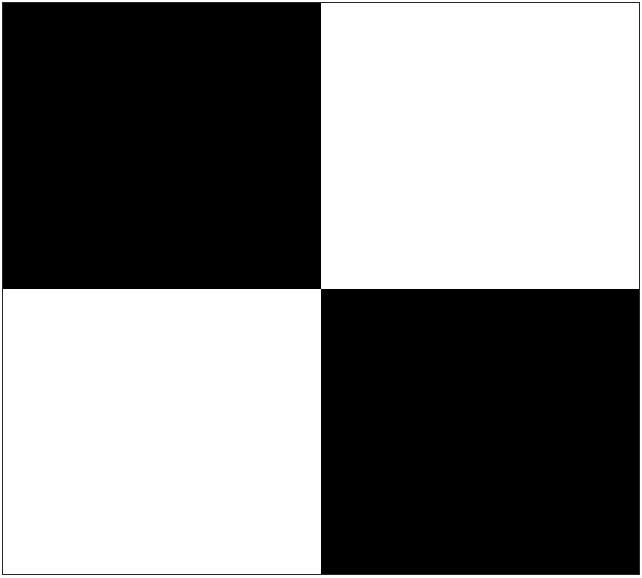}
				\caption{}
    \end{subfigure}%
		\hfill 
		\begin{subfigure}[t]{0.2\textwidth}
        \includegraphics[width=0.95\linewidth]{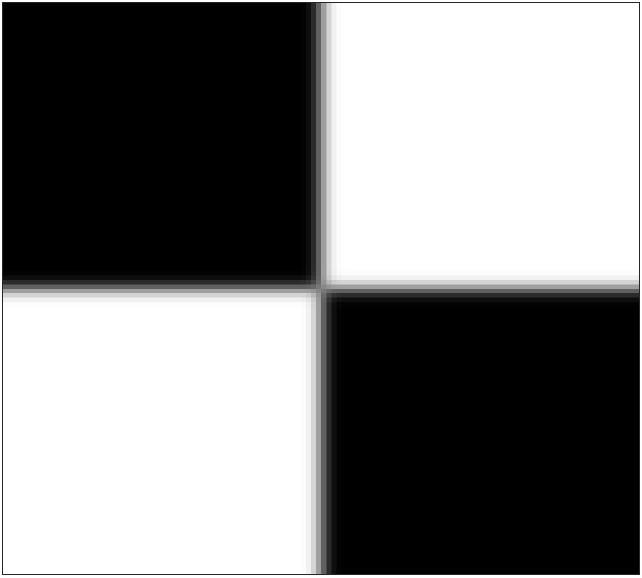}
				\caption{}
    \end{subfigure}%
		\hfill 
    \begin{subfigure}[t]{0.2\textwidth}
        \includegraphics[width=0.95\linewidth]{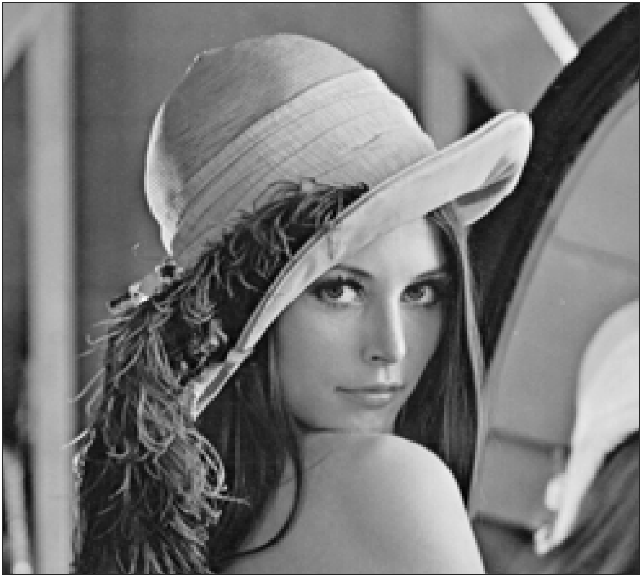}
				\caption{}
    \end{subfigure}%
		\hfill 
		\begin{subfigure}[t]{0.2\textwidth}
        \includegraphics[width=0.95\linewidth]{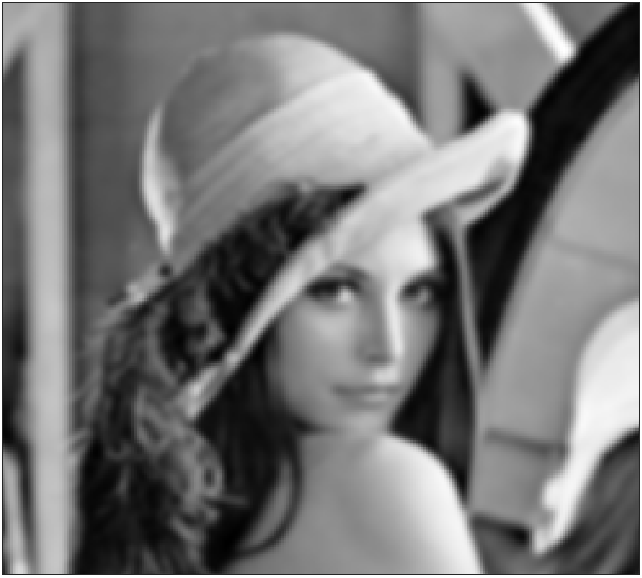}
				\caption{}
    \end{subfigure}%
		\hfill
		\begin{subfigure}[t]{0.2\textwidth}
				\includegraphics[width=0.95\linewidth]{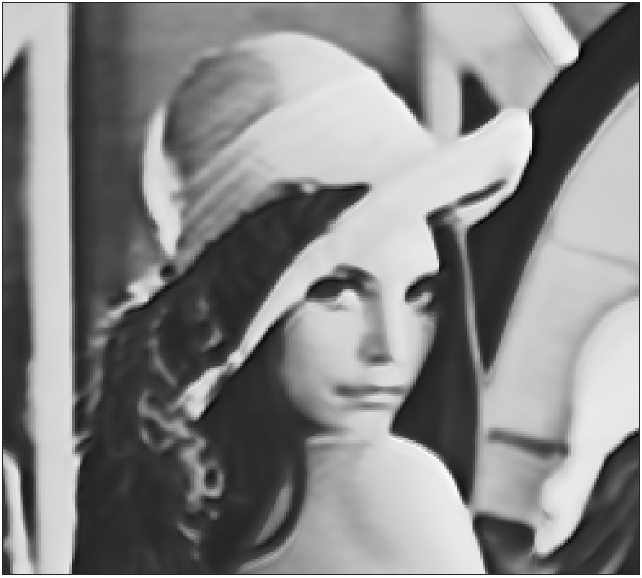}
				\caption{}
    \end{subfigure}
	
    \begin{subfigure}[t]{0.2\textwidth}
        \includegraphics[width=0.95\linewidth]{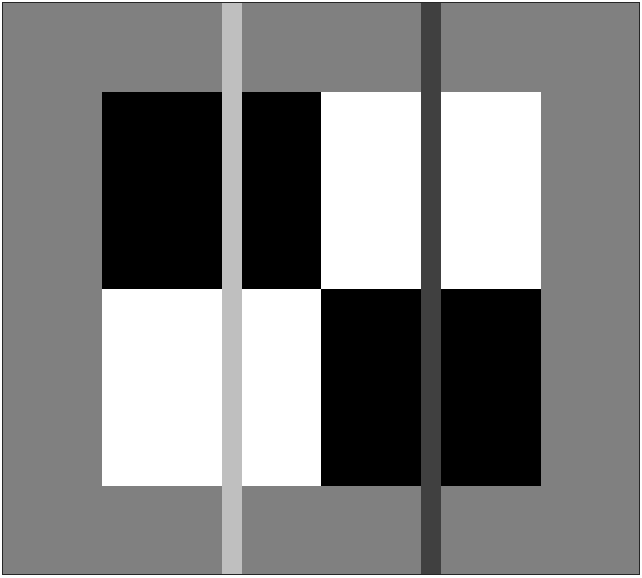}
				\caption{}
    \end{subfigure}%
		\hfill 
		\begin{subfigure}[t]{0.2\textwidth}
        \includegraphics[width=0.95\linewidth]{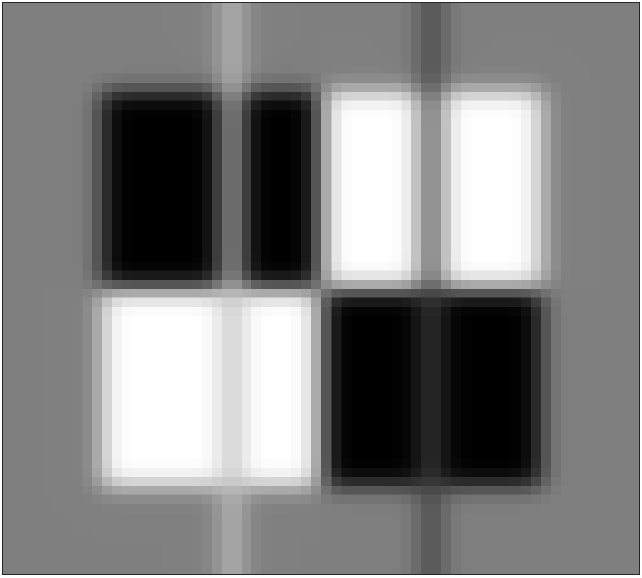}
				\caption{}
    \end{subfigure}%
		\hfill 
    \begin{subfigure}[t]{0.2\textwidth}
        \includegraphics[width=0.95\linewidth]{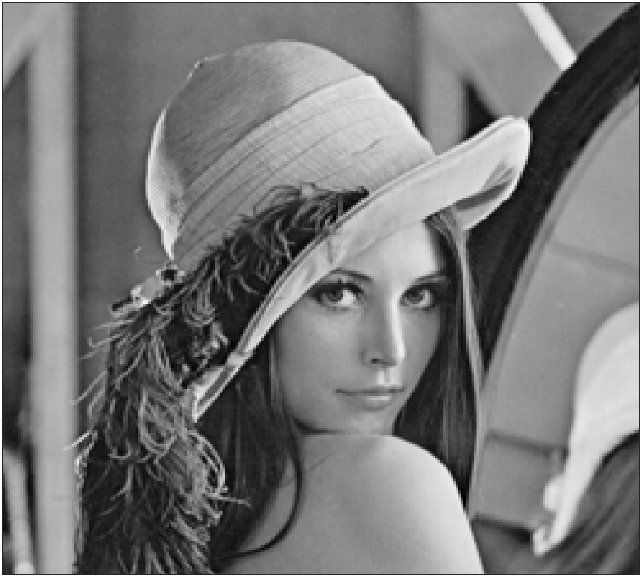}
				\caption{}
    \end{subfigure}%
		\hfill 
		\begin{subfigure}[t]{0.2\textwidth}
        \includegraphics[width=0.95\linewidth]{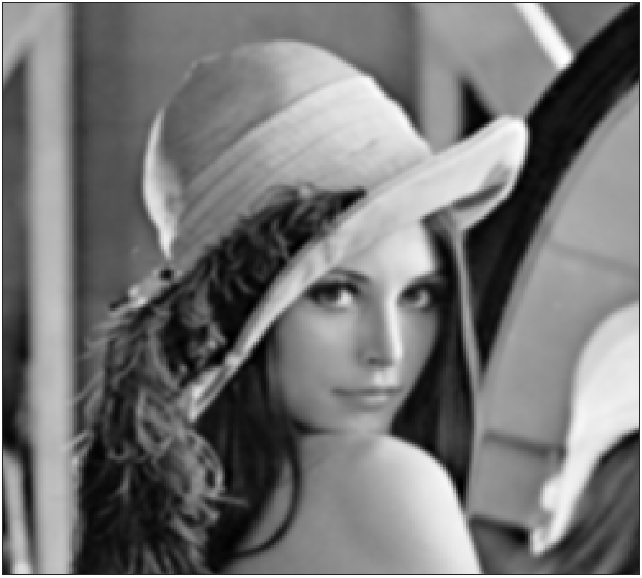}
				\caption{}
    \end{subfigure}%
		\hfill
		\begin{subfigure}[t]{0.2\textwidth}
				\includegraphics[width=0.95\linewidth]{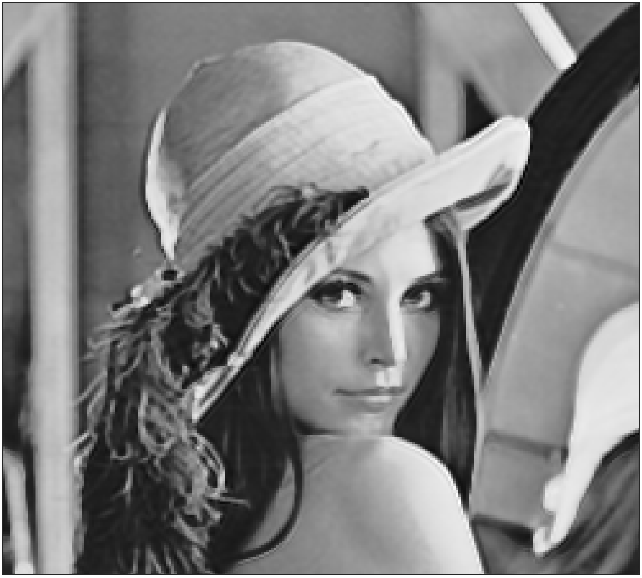}
				\caption{}
    \end{subfigure}%
		
    \begin{subfigure}[t]{0.2\textwidth}
        \includegraphics[width=0.95\linewidth]{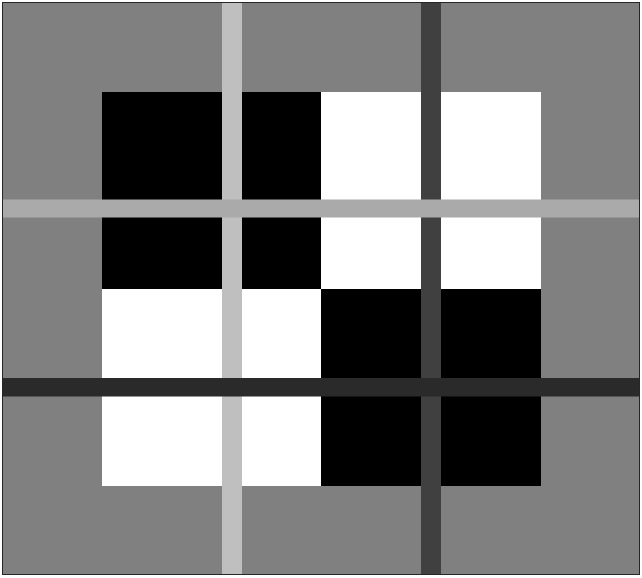}
				\caption{}
    \end{subfigure}%
		\hfill 
		\begin{subfigure}[t]{0.2\textwidth}
        \includegraphics[width=0.95\linewidth]{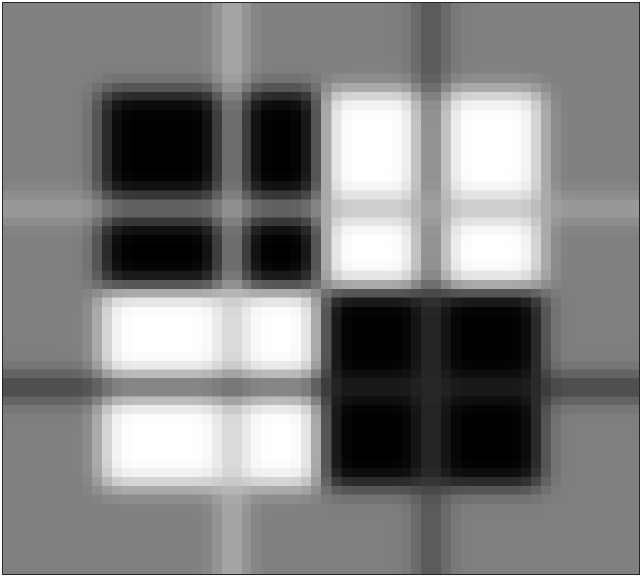}
				\caption{}
    \end{subfigure}%
		\hfill 
    \begin{subfigure}[t]{0.2\textwidth}
        \includegraphics[width=0.95\linewidth]{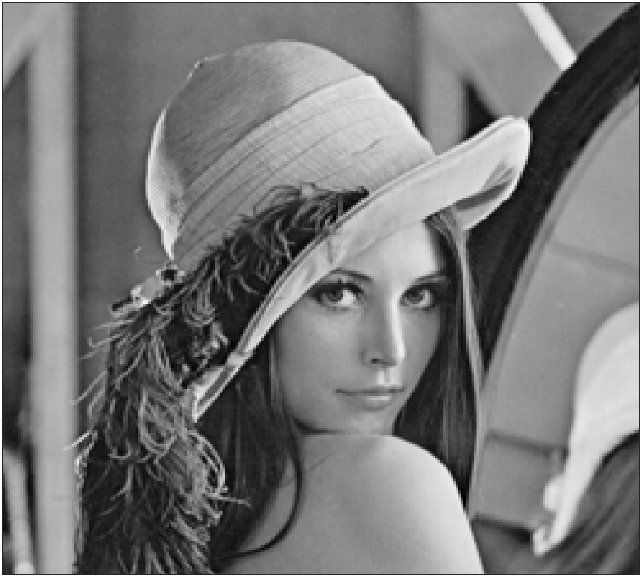}
				\caption{}
    \end{subfigure}%
		\hfill 
		\begin{subfigure}[t]{0.2\textwidth}
        \includegraphics[width=0.95\linewidth]{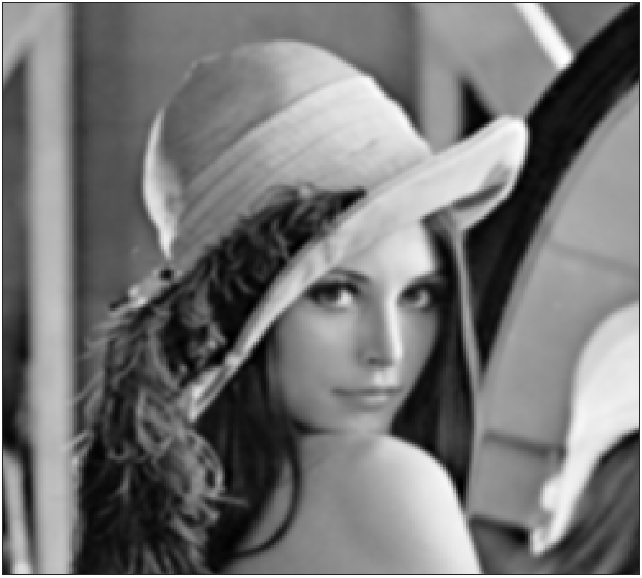}
				\caption{}
    \end{subfigure}%
		\hfill
		\begin{subfigure}[t]{0.2\textwidth}
				\includegraphics[width=0.95\linewidth]{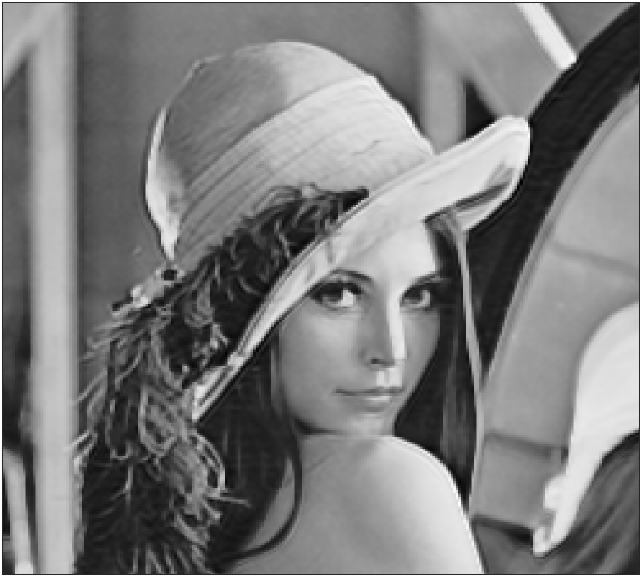}
				\caption{}
    \end{subfigure}%
		\\
\caption{{\small Image deblurring experiments: Columns 1-2 present the training image pair;
columns 3-4 present ground truth and input observation, column 5 present RNN estimation. First line - Gaussian blur with $\sigma=1.6$. Lines 2-3 Gaussian blur with $\sigma=1.2$.}}
\label{fig_queen} 
\end{figure*}

To further investigate the potential of the proposed framework we trained the RNN framework with the chessboard example presented in~\ref{fig_queen}(a)-(b). The obtained results for Lena image are presented in figures~\ref{fig_queen}(c)-(e).
As can be seen, surprisingly, the network is able to learn and generalize a style transfer with a relatively degenerated training example. Additional examples of simplified synthetic training pairs and the corresponding results are presented in Fig~\ref{fig_queen}(f)-(o). These experiments give rise to potential deeper theoretical investigation into the fundamental understanding of the pattens learned by the model and further possibilities of improvement in generalization in image-to-image translation tasks.

\begin{table*}[h!]
\begin{center}
\caption{Comparison of running time for different image restoration methods}
\label{Table 3}
{\small
	\begin{tabular}{ |m{12em}|m{8em}|m{10em}|}
    \hline
																		& Runtime per image  		& Computational Resource  				\\ \hline \hline
									 
Zero-shot super resolution	SRx2
\cite{Shocher:2018}									&	1-5 minutes
																	
																															& Tesla K-80  GPU  (24GB GDDR5 memory)										\\ \hline

Deep image prior \cite{Ulyanov:2018}
(reported in \cite{Mataev:2019})    & 6.6 minutes      				& GeForce RTX 2080 Ti GPU (12GB GDDR6 memory) 

\\ 
\cline{1-2}
											
DeepRED \cite{Mataev:2019} 
																		& 9.5 minutes     				& 
\\ \hline
    
RNN - training (ours)								& 15-30 seconds			& laptop GPU NVIDIA GeForce GTX Ti 1650 (4GB GDDR6 memory)  							\\
RNN - inference (ours)							& 500 milliseconds	& 
\\ 
			\hline 

UNET-GAN - training (ours)			& 4.76 minutes					& laptop GPU NVIDIA RTX 3500 (12GB GDDR6 memory)  							\\
UNET-GAN - inference (ours)			& 80 milliseconds			& 
\\ 
			\hline 
  \end{tabular} 
} \\
\end{center}
\end{table*}

\paragraph{Runtime}
One of the main advantages of the proposed approach is substantial training and inference speed. Tensorflow \cite{TF:2016} implementation converges in an average of 14.29 seconds on a laptop GPU (NVIDIA GeForce GTX Ti 1650 with 4GB video memory) and 2.01 minutes on i-7 CPU. Pytorch \cite{Pytorch:2019} implementation training converges on average in 30.73 seconds on a GPU NVIDIA GeForce RTX 2080 Ti, and 2.73 minutes on i-7 CPU. RNN inference is about 500 msec. U-Net inference is less than 100 msec. For reference, RED-SD takes 15-20 minutes for each image, on a CPU. Using internal learning, Zero-Shot-Single Image Super-Resolution (ZSSR) \cite{Shocher:2018}, the neural net is retrained for each image. The authors state that although training is done at test time, the average runtime for SRx2 is only 9 sec on Tesla V100 GPU or 54 sec on K-80 (average taken on BSD100 dataset), which increases to $~$5 minutes, depending on the desired resolution. 
Both are relatively powerful GPUs. 
Deep image prior \cite{Ulyanov:2018} and DeepRED \cite{Mataev:2019} require retraining for several minutes for each new test image. Table~\ref{Table 3} compares training and testing times for different methods. 
Our framework therefore introduces significantly efficient training and testing, since it requires training only once at a cost of less than a minute up to few minutes, depending on the selected architecture, and inference time is of the order of tens of milliseconds. 
This advantage could be leveraged for applications that require real-time training, as well as for research purposes. 


\section{System Mismatch}\label{sec6}

\begin{figure*}[t]
    \begin{subfigure}[t]{0.14\textwidth}
        \includegraphics[width=0.99\linewidth]{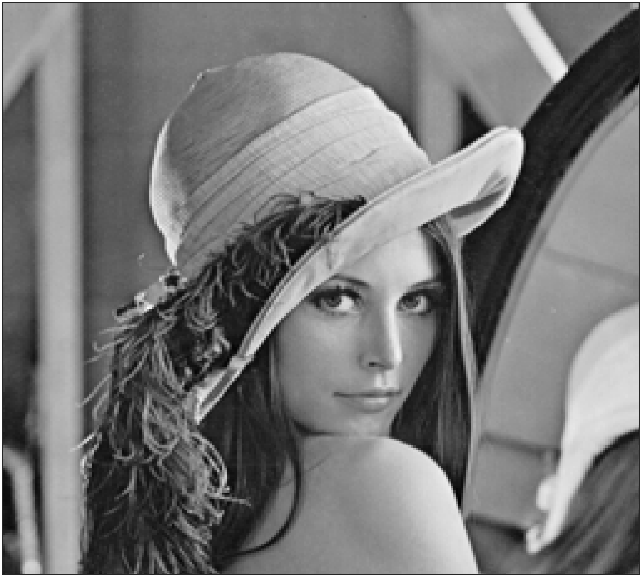}
				\caption{}
    \end{subfigure}%
		\hfill 
		\begin{subfigure}[t]{0.14\textwidth}
        \includegraphics[width=0.99\linewidth]{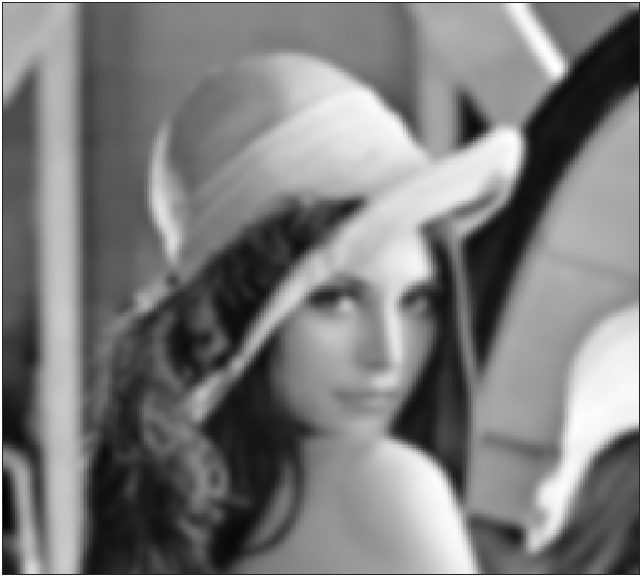}
				\caption{}
    \end{subfigure}%
		\hfill 
    \begin{subfigure}[t]{0.14\textwidth}
        \includegraphics[width=0.99\linewidth]{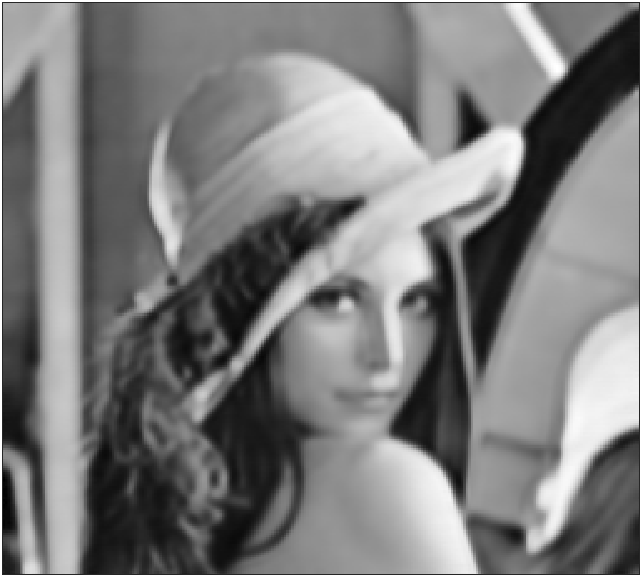}
				\caption{}
    \end{subfigure}%
		\hfill 
		\begin{subfigure}[t]{0.14\textwidth}
        \includegraphics[width=0.99\linewidth]{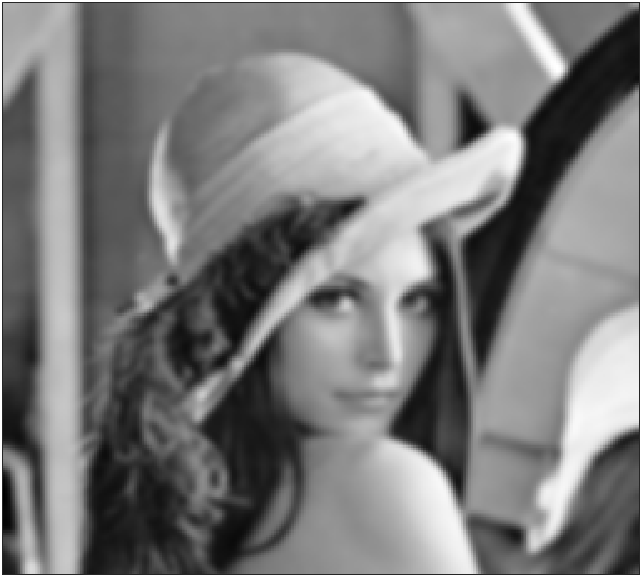}
				\caption{}
    \end{subfigure}%
		\hfill
		\begin{subfigure}[t]{0.14\textwidth}
        \includegraphics[width=0.99\linewidth]{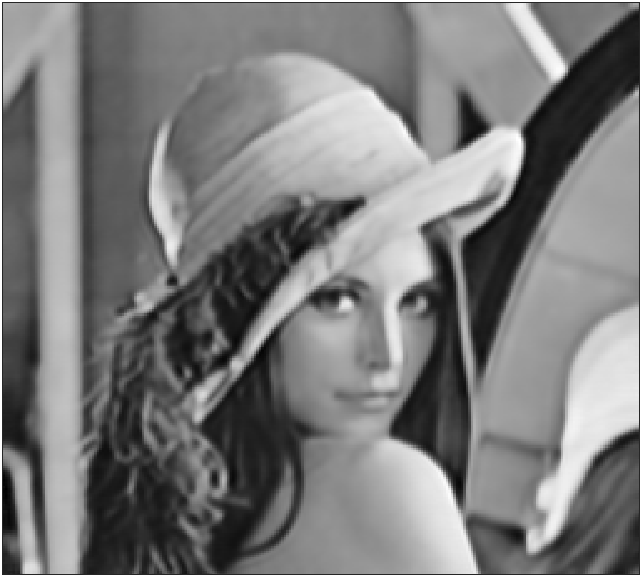}
				\caption{}
    \end{subfigure}%
		\hfill
		\begin{subfigure}[t]{0.14\textwidth}
        \includegraphics[width=0.99\linewidth]{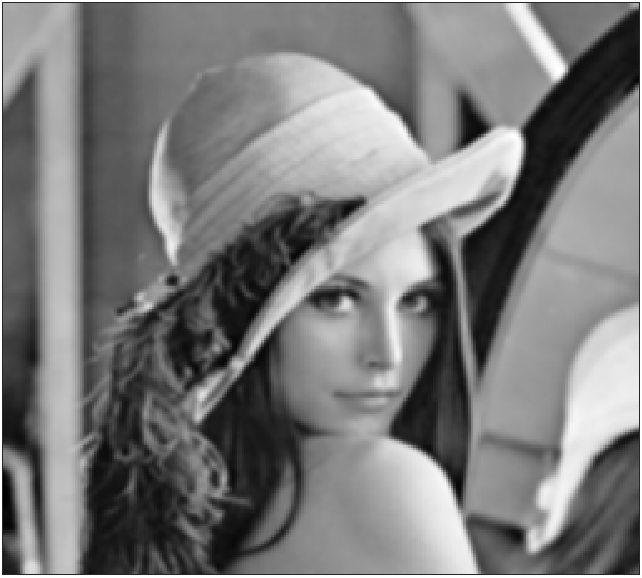}
				\caption{}
    \end{subfigure}%
		\hfill
		\begin{subfigure}[t]{0.14\textwidth}
        \includegraphics[width=0.99\linewidth]{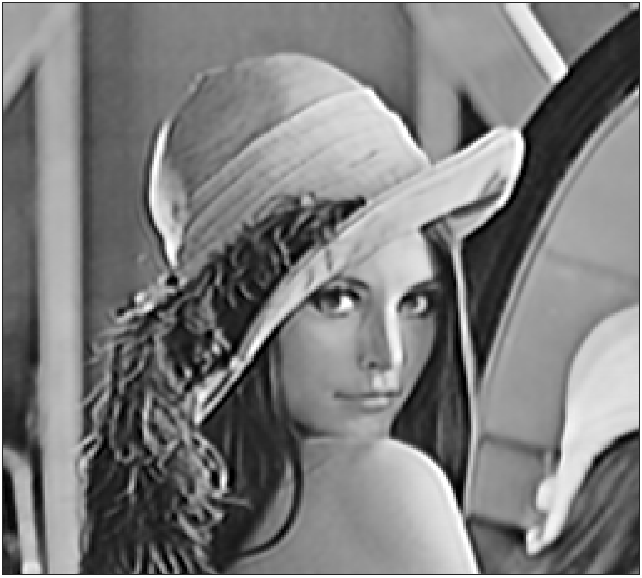}
				\caption{}
    \end{subfigure}%
\caption{{\small Deblurring Lena for varying Gaussian blurring kernels: (a) Ground truth; Blurred image and reconstruction, for Gaussian filter with: (b)-(c) $\sigma_t=2$; (d)-(e) $\sigma_t=1.6$;
(f)-(g) $\sigma_t=1$. RNN-GAN trained with the image boat and Gaussian blur with $\sigma_s=2$.}}
\label{fig1}
\end{figure*}

\begin{figure}[h!]
\centering
\includegraphics[width=4.5cm, height=3.25cm]{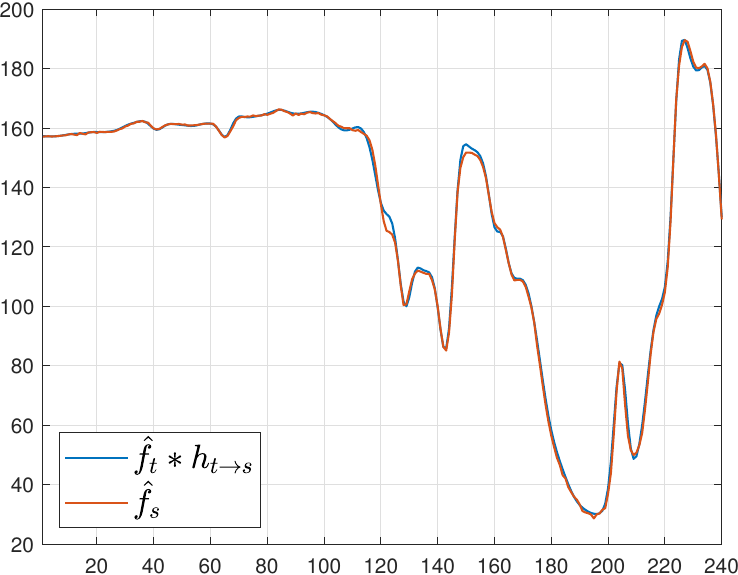}
\caption{1D plot of column 55 of image boat prediction.}
\label{fig8}
\vspace{-15pt}
\end{figure}

Assume an observed blurred image $g_s[m,n]=f[m,n]*h_s[m,n]$, from a source domain $\mathcal{S}$ as ground truth 
where $f[m,n]$ is the source original image and $h_s[m,n]$ is a Gaussian filter with standard deviation $\sigma_s$, and $*$ denotes the convolution operator.
The learning system $\mathcal{F}_s :g_s \rightarrow f$ is trained to invert an image blurred with $h_s[m,n]$.
Now, assume we are given a target image $g_t[m,n]=f[m,n]*h_t[m,n]$, from a source domain $\mathcal{T}$, where $h_t[m,n]$ is a Gaussian filter with standard deviation $\sigma_t$, and we are trying to deblur $g_t$ using our trained predictor $\mathcal{F}_s$.  
What should we expect?
For a Gaussian PSF we know the convolution of two Gaussians with mean $\mu_1, \mu_2$ and variance $\sigma^2_1$, $\sigma^2_2$ is a Gaussian with mean $\mu=\mu_1+\mu_2$ and variance $\sigma^2 =\sigma^2_1+\sigma^2_2$.
Therefore, 
\begin{enumerate}
\item If $\sigma_s \leq \sigma_t$, then $g_t= f*h_t = f * h_s * h_{s\rightarrow t}$, where $h_{s \rightarrow t}$ is a Gaussian filter with $\sigma^2_{s \rightarrow t}=\sigma_t^2-\sigma_s^2$. In this case applying our trained predictor yields
\begin{equation}
\mathcal{F}_s \big(g_t[m,n]\big) =\hat{f}[m,n]\Big|_{\hat{f}*h_s=g_t} = f*h_{s \rightarrow t}.
\end{equation}
In other words, the estimated image $\hat{f}$ is a blurred version of $f$, by the ``residual blur''.\\
\item If $\sigma_s > \sigma_t$, then $g_s=f*h_t*h_{t \rightarrow s}$, where $h_{t \rightarrow s}$ is a Gaussian filter with $\sigma^2_{t \rightarrow s}=\sigma_s^2-\sigma_t^2$. In this case applying our trained predictor yields
\begin{equation}
\mathcal{F}_s \big(g_t[m,n]\big) =\hat{f}[m,n]\Big|_{\hat{f}*h_s=g_t} =
\hat{f}[m,n]\Big|_{\hat{f}*h_{t \rightarrow s}=f}.
\end{equation}
The prediction is a deblurred version of the original image. 
\end{enumerate}
To test this analysis, we trained the proposed RNN-GAN with the image boats blurred with $\sigma_s=2$.
Figure~\ref{fig1} shows an example of the deblurred Lena for Gaussian blur of $\sigma_t=\{2,1.6,1\}$. 
Figure~\ref{fig8} compares a 1D plot of column 55 in $\hat{f}_s$  - the restored image boats with input blur with $\sigma_s=2$, and the corresponding line in $\hat{f}_t$ - the restored image with input blur $\sigma_t=1$ convolved with $h_{t \rightarrow s}$, with $\sigma_{t \rightarrow s}=\sqrt{3}$.
As can be seen the above analysis aligns with these empirical results.

\section{Conclusions}
We investigated a patch-based supervised learning approach based on a single-image example, and provided a proof-of-concept for its applicability to image-to-image restoration.
We challenged the common assumption that image-to-image learning is achievable by either conventional supervised-learning with large datasets or via internal-learning. 
Future work can further explore the proposed approach for image restoration with varying distortions, motion blur, and for super-resolution of real low resolution images, as well as expand the proposed approach to other tasks, such as segmentation. 
Our proposed point-of-view could be easily derived for general signals, thus applicable to other modalities. 

\appendix

\section{Sparse Representations}\label{A.1}
Sparse coding (SC) is a popular task in many fields, such as: image processing \cite{Elad}, computer vision \cite{Jarrett:2009},  compressed sensing \cite{Donoho}, ultrasound imaging \cite{Bendory:2016}, seismology \cite{Pereg:2017A,Pereg:2017b,Pereg:2019A,Pereg:2021}, and visual neurosciense \cite{Olshausen:1996,Lee:2009}. 
A sparse representations model \cite{Elad} assumes a signal $\mathbf{y}\in\mathbb{R}^{N \times 1}$ that is analyzed as a sparse linear combination of some dictionary basis components:
\begin{equation} \label{A1}
\mathbf{y}=\mathbf{D}\mathbf{z},
\end{equation}
where $\mathbf{D}\in\mathbb{R}^{N \times M}$ is a matrix called the dictionary, built of the atoms $\mathbf{d}_i \in\mathbb{R}^{N \times 1}, \ i=1,...,M$, as its columns, and
$\mathbf{z} \in \mathbb{R}^{M \times 1}$ is the sparse vector of the atoms weights. 
Sparse coding, that is, the recovery of $\mathbf{z}$, has been the center of abundant research efforts. 
Finding the sparsest solution, the one with the smallest $\ell_0$-norm, is basically attempting to solve
\begin{equation}\label{A2}
(P_0): \qquad \min_\mathbf{z} \|\mathbf{z}\|_0 \qquad 	\mathrm{s.t.}	 \qquad 	\mathbf{y}=\mathbf{D}\mathbf{z},
\end{equation}
where $\|\mathbf{z}\|_0$ denotes the number of non-zeros in $\mathbf{z}$.
Unfortunately, $P_0$ is in general NP-Hard \cite{Natarajan:1995}, therefore the $\ell_0$-norm is often replaced with the $\ell_1$-norm
 \begin{equation}\label{A3}
(P_1): \qquad \min_\mathbf{z} \|\mathbf{z}\|_1 \qquad 	\mathrm{s.t.}	 \qquad 	\mathbf{y}=\mathbf{D}\mathbf{z},
\end{equation}
where $\|\mathbf{z}\|_1 \triangleq \sum_i |z_i|$.
In many real-life scenarios, such as in the presence of noise or when some error is allowed, we solve 
\begin{equation}\label{A4}
(P_{1,\varepsilon}): \qquad \min_\mathbf{z} \|\mathbf{z}\|_1 \qquad 	\mathrm{s.t.}	 \qquad 	 \|\mathbf{y}-\mathbf{D}\mathbf{z}\|_2 \leq \varepsilon,
\end{equation}
where $\|\mathbf{z}\|_2 \triangleq \sqrt{\sum_i z^2_i}$.
The sparsest solution to $P_0$ and $P_1$ is unique under certain conditions, and can be obtained with known algorithms, such as orthonormal matching pursuit (OMP) or basis pursuit (BP), depending on the dictionary's properties and the degree of sparsity of $\mathbf{z}$. That is, when $\| \mathbf{z} \|_0 < \frac{1}{2}\Big( 1+ \frac{1}{\mu(\mathbf{D})} \Big)$,
where $\mu(\mathbf{D})$ is the mutual coherence defined as
\begin{equation}\label{A5}
\mu(\mathbf{D})=\max_{i \neq j} \frac{\Big|\mathbf{d}^T_i\mathbf{d}_j\Big|}{\|\mathbf{d}_i\|_2 \cdot \|\mathbf{d}_j\|_2},
\end{equation}
the true sparse code $\mathbf{z}$ can be perfectly recovered \cite{Candes:2011}.

An intuitive way to recover $\mathbf{z}$ is to project $\mathbf{y}$ on the dictionary, and then extract the atoms with the strongest response by taking a hard or a soft threshold, i.e.,
$\mathbf{z}=\mathcal{H}_\beta(\mathbf{D}^T\mathbf{y})$ or $\mathbf{z}=\mathcal{S}_\beta(\mathbf{D}^T\mathbf{y})$,
where the hard threshold and the soft threshold operators are respectively defined as
\begin{center}
\begin{tabular}{ccc}
$
\mathcal{H}_\beta(z)=  
\begin{cases}
z, & |z| > \beta \\
0, & |z| \leq \beta\\
\end{cases},
$
&
and 
&
$
\mathcal{S}_\beta(z)=  
\begin{cases}
z+\beta, & z<-\beta \\
0, &  |z| \leq \beta\\
z-\beta, & z>\beta
\end{cases}.
$
\end{tabular}
\end{center}

Note that the ReLU activation function obeys
\begin{equation*}
\mathrm{ReLU}(z-\beta)= \max (z-\beta,0) =\mathcal{S}^{+}_\beta(z) \triangleq 
\begin{cases}
0, &  z \leq \beta\\
z-\beta, & z>\beta
\end{cases}.
\end{equation*}
Therefore, the soft threshold solution can be also written as
\begin{equation*}
\mathbf{z}
=\mathcal{S}^{+}_\beta(\mathbf{D}^T\mathbf{y})-\mathcal{S}^{+}_\beta(-\mathbf{D}^T\mathbf{y})
= \mathrm{ReLU}(\mathbf{D}^T\mathbf{y}-\beta)-\mathrm{ReLU}(-\mathbf{D}^T\mathbf{y}-\beta).
\end{equation*}
It is possible to assume a nonnegative sparse code such that the weights are solely positive, essentially assuming a compounded dictionary $[\mathbf{D},-\mathbf{D}]$ \cite{Papyan:2017}. Hence a nonnegative model does not affect the expressiveness of the model.
Perfect support recovery by simple thresholding is guaranteed only when $\| \mathbf{z} \|_0 < \frac{1}{2} \Big( 1+ \frac{1}{\mu(\mathbf{D})} \frac{|\mathbf{z}|_{\mathrm{min}}}{|\mathbf{z}|_{\mathrm{max}}} \Big)$, where $|\mathbf{z}|_{\mathrm{min}}$, and $|\mathbf{z}|_{\mathrm{max}}$ are the minimum and maximum values of the vector $|\mathbf{z}|$ on the support, implying that this approach may have stability issues when the data is unbalanced.

In the special case where $\mathbf{D}$ is a convolutional dictionary, the task of extracting $\mathbf{z}$ is referred to as convolutional sparse coding (CSC). In this case, the dictionary $\mathbf{D}$ is a convolutional matrix constructed by shifting a local matrix of filters in all possible positions. 
The forward pass of CNNs is equivalent to the layered thresholding algorithm designed to solve the CSC problem \cite{Papyan:2017}.

\bibliographystyle{plain}

\bibliography{dpereg_iccv_2023}

\end{document}